\documentclass[aps,prc,superscriptaddress,showpacs,twocolumn,floatfix]{revtex4}

\usepackage{graphicx}

\def\gtorder{\mathrel{\raise.3ex\hbox{$>$}\mkern-14mu
 \lower0.6ex\hbox{$\sim$}}}
\def\ltorder{\mathrel{\raise.3ex\hbox{$<$}\mkern-14mu
 \lower0.6ex\hbox{$\sim$}}}

\def\mugegm{\mu_p G_E^p / G_M^p}

\def\etal{\emph{et al.}}

\begin{document}

\title{Flavor decomposition of the nucleon electromagnetic form factors at low $Q^2$}

\author{I.~A.~Qattan}
\affiliation{Khalifa University of Science, Technology and Research, Department
of Applied Mathematics and Sciences, P.O. Box 127788, Abu Dhabi, UAE}

\author{J.~Arrington}
\affiliation{Physics Division, Argonne National Laboratory, Argonne, Illinois,
60439, USA}

\author{A.~Alsaad}
\affiliation{Jordan University of Science and Technology, Department of
Physical Sciences, P.O. Box 3030, Irbid 22110, Jordan}

\date{\today} 


\begin{abstract}
\begin{description}

\item[Background] The spatial distribution of charge and magnetization within the proton is encoded
in the elastic form factors. These have been precisely measured in elastic electron scattering, and
the combination of proton and neutron form factors allows for the separation of the up- and down-quark
contributions.

\item[Purpose] 
In this work, we extract the proton and neutron form factors from world's data with an emphasis on
precise new data covering the low-momentum region, which is sensitive to the large-scale structure of
the nucleon. From these, we separate the up- and down-quark contributions to the proton form factors.

\item[Method] 
We combine cross section and polarization measurements of elastic electron-proton scattering to 
separate the proton form factors and two-photon exchange (TPE) contributions. We combine the proton
form factors with parameterization of the neutron form factor data and uncertainties to separate the
up- and down-quark contributions to the proton's charge and magnetic form factors.

\item[Results]
The extracted TPE corrections are compared to previous phenomenological extractions, TPE calculations,
and direct measurements from the comparison of electron and positron scattering. The flavor-separated
form factors are extracted and compared to models of the nucleon structure.

\item[Conclusions]
With the inclusion of the precise new data, the extracted TPE contributions show a clear change of
sign at low $Q^2$, necessary to explain the high-$Q^2$ form factor discrepancy while being consistent
with the known $Q^2 \to 0$ limit.  We find that the new Mainz data yield a significantly different
result for the proton magnetic form factor and its flavor-separated contributions.  We also observe
that the RMS radius of both the up- and down-quark distributions are smaller than the RMS charge radius of
the proton.

\end{description}
\end{abstract}

\pacs{25.30.Bf, 13.40.Gp, 14.20.Dh}

\maketitle

\section{Introduction}

The electromagnetic form factors of the proton and neutron, $G_E^{(p,n)}$ and $G_M^{(p,n)}$,
are fundamental quantities which provide information on the spatial distributions of charge
and magnetization in nucleons. The form factors are measured using electron scattering where
the incident electron scatters from a nucleon target through the exchange of a virtual photon
which serves as the sole mediator of the electron-nucleon electromagnetic interaction. By
increasing $Q^2$, the four-momentum squared of the virtual photon, the virtual photon becomes
more sensitive to the small scale internal structure of the nucleon. 

In electron scattering there are primarily two methods used to extract the
proton form factors. The first is the Rosenbluth or Longitudinal-Transverse
(LT) separation method~\cite{rosenbluth50}, which uses measurements of the
unpolarized cross section, and the second is the polarization transfer or
polarized target (PT) method~\cite{dombey69}, which requires measurement of
the spin-dependent cross section. A significant difference is observed between
LT and PT extractions of the proton form factors~\cite{arrington07a, perdrisat07},
which is currently believed to be the results of larger-than-expected two-photon 
exchange (TPE) contributions~\cite{carlson07, arrington11b}.

In this paper, we extract the proton form factors from a combined analysis of LT and PT
measurements, accounting for TPE contributions following the approach of Ref.~\cite{qattan11a},
but with an emphasis on recent low-$Q^2$ data. The combined analysis allows us to extract the
TPE contribution to the e-p elastic cross section, which we compare to other phenomenological 
extractions and to direct calculations of TPE effects meant to be valid in the lower $Q^2$ regime,
as well as to recent direct measurements of TPE contributions from the comparison of electron-proton
and positron-proton elastic scattering~\cite{adikaram14, rachek14}.

With the inclusion of neutron form factor measurements, we separate the nucleon form factors into
their up-quark and down-quark contributions~\cite{cates11, qattan12}. New cross section and polarization
measurements at low $Q^2$ allow for a more detailed examination of the flavor-separated form factors in
the region sensitive to the large-distance behavior of the proton charge and magnetization
distributions. We compare these results to previous extractions, as well as to models of the up- and
down-quark contributions to the proton form factors.

\section{Overview of the Methods}

\subsection{Rosenbluth Separation Method}

In the Rosenbluth separation method, the reduced cross section $\sigma_{R}$ for elastic e-p scattering
in the Born or one-photon exchange (OPE) approximation is:
\begin{equation} \label{eq:reduced2}
\sigma_{R} = \big(G_M^p(Q^2)\big)^2 + \frac{\varepsilon}{\tau} \big(G_E^p(Q^2)\big)^2,
\end{equation}
where $\tau = Q^2/4M_{p}^2$, $M_{p}$ is the mass of the proton, and
$\varepsilon$ is the virtual photon longitudinal polarization parameter,
defined as $\varepsilon^{-1} = \big[ 1 + 2 (1+\tau)
\tan^2({\frac{\theta_e}{2}})\big]$, where $\theta_e$ is the scattering
angle of the electron. For a fixed $Q^2$ value, the reduced cross section
$\sigma_{R}$ is measured at several $\varepsilon$ points, and a linear fit of
$\sigma_{R}$ to $\varepsilon$ gives $(G_M^p)^2$ as the intercept and
$(G_E^p)^2/\tau$ as the slope.

By assuming isospin and charge symmetry and neglecting strange quarks contribution,
the nucleon form factors can be expressed in terms of the up- and down-quark contributions~\cite{miller90,beck01}
\begin{eqnarray} \label{eq:FFs_Flavor1}
G_{E,M}^p = \frac{2}{3} G_{E,M}^u - \frac{1}{3} G_{E,M}^d,\nonumber \\
G_{E,M}^n = \frac{2}{3} G_{E,M}^d - \frac{1}{3} G_{E,M}^u,~
\end{eqnarray}
where in this convention $G_{E,M}^u$ represents the contribution from the up-quark distribution in the
proton and the down-quark distribution in the neutron.

Solving for $G_{E,M}^u$ and $G_{E,M}^d$ from Eq.~(\ref{eq:FFs_Flavor1}) above, we get the following
expression for the up- and down-quark contribution to the proton form factors
\begin{eqnarray} \label{eq:FFs_Flavor2}
G_{E,M}^u = 2 G_{E,M}^p + G_{E,M}^n,~~~ G_{E,M}^d = G_{E,M}^p + 2 G_{E,M}^n.
\end{eqnarray}
In the limit $Q^2 = 0$, this yields $G_E^u = 2, G_E^d = 1, G_M^u = (2\mu_{p} + \mu_n) = 
3.67\mu_N,$ and $G_M^d = (\mu_{p} + 2\mu_n) = -1.03\mu_N$, where $\mu_N$ is the 
nuclear magneton.

\subsection{Recoil Polarization Measurements}

In the recoil polarization method, a beam of longitudinally polarized
electrons scatters elastically from unpolarized proton target. The electrons
transfer their polarization to the unpolarized protons. By simultaneously
measuring the transverse, $P_{t}$, and longitudinal, $P_{l}$, polarization
components of the recoil proton, one can determine the ratio
$\mu_{p}G_E^p/G_M^p$ in the OPE~\cite{dombey69, akhiezer74, arnold81}:
\begin{equation} \label{eq:ratio}
R = \frac{\mu_p G_E^p}{G_M^p} = -\frac{P_t}{P_l} \frac{(E+E')}{2M_{p}} \tan({\frac{\theta_e}{2}}),
\end{equation}
where $E$ and $E'$ are the initial and final energy of the incident electron,
respectively.  The ratio can be extracted in a similar fashion using polarized
beams and targets by measuring the asymmetry for two different spin
directions~\cite{arrington07a,perdrisat07}.

The two methods yield strikingly different results, with values of $\mu_{p}G_E^p/G_M^p$
differing almost by a factor of three at high $Q^2$. In the LT separation
method, the ratio shows approximate form factor scaling, $\mu_p G_E^p/G_M^p
\approx 1$, albeit with large uncertainties at high $Q^2$ values. The recoil
polarization method yields a ratio that decreases roughly
linearly with increasing $Q^2$, with some hint of flattening out above 5 (GeV/c)$^2$.

\subsection{Two-Photon Exchange Contributions} \label{sec:TPE}

To reconcile these measurements, several studies suggested that missing higher order radiative corrections to
the electron-proton elastic scattering cross sections, in particular two-photon exchange (TPE) diagrams,
may explain the discrepancy~\cite{guichon03, arrington03a, arrington04a}. The role of TPE effects was
studied extensively both theoretically~\cite{blunden03, blunden05a, kondratyuk05, chen04, borisyuk06a,
borisyuk06b} and phenomenologically~\cite{guichon03, tvaskis06, borisyuk06c, chen07, arrington05,
guttmann11, qattan11b, borisyuk07, qattanphd}. Most calculations suggested that the TPE corrections
are relatively small, but have a significant angular dependence which mimics the effect of a larger
value of $G_E^p$. Detailed reviews of the role of the TPE effect in electron-proton scattering can be
found in~\cite{arrington11b, carlson07}.

Experimentally, several measurements were performed to verify the discrepancy~\cite{christy04, qattan05}
and to try and measure or constrain TPE contributions. Precise examinations of the $\varepsilon$
dependence of $\sigma_{R}$~\cite{qattanphd, tvaskis06, chen07} found no deviation from the linear
behavior predicted in the OPE approximation. Another measurement was performed to look for TPE
effects by extracting $\mugegm$ at fixed $Q^2$ as a function of scattering angle~\cite{meziane11}. In
the Born approximation, the result should be independent of scattering angle, and no deviation from the
OPE prediction was observed.

Based on the observations above, it is possible to try and extract the TPE contributions based on the
observed discrepancy between the LT and PT results. Assuming that the TPE contributions are linear in
$\varepsilon$ and that the PT results do not depend on $\varepsilon$, and knowing that the TPE
contribution must vanish in the forward limit ($\varepsilon \to 1$)~\cite{rekalo04, chen07}, it is possible to
extract the TPE contribution to the unpolarized cross section in a combined analysis of LT and PT
data~\cite{guichon03, borisyuk06c, chen07, arrington05, qattan11a}. Where polarization data exist as a
function of $\varepsilon$, it is possible to attempt to extract the TPE amplitudes with fewer
assumptions~\cite{guttmann11, borisyuk11}, though with relatively large uncertainties.

The most direct technique for measuring TPE is the comparison of electron-proton and positron-proton
scattering. The leading TPE contribution comes from the interference of the OPE and TPE amplitudes, and so
has the opposite sign for positron and electron scattering. The only other first-order radiative correction 
which depends on the lepton sign is the interference between diagrams with Bremsstrahlung from the electron
and proton, and this contribution is generally small. Thus, after correcting the measured ratio for the
Bremsstrahlung interference term, the comparison of positron and electron scattering allows for the most
direct measurement of  TPE contributions.

The ratio $R_{e^{+} e^{-}}(\varepsilon,Q^2)$
is defined as
\begin{equation} \label{eq:ratiopostelect}
R^{\mbox{raw}}_{e^{+} e^{-}}(\varepsilon,Q^2) = \frac{\sigma(e^{+}p~\rightarrow e^{+}p)}
{\sigma(e^{-}p~\rightarrow e^{-}p)}. 
\end{equation}  
After correcting for the electron-proton Bremsstrahlung interference term and the conventional 
charge-independent radiative corrections, the cross section ratio $R_{e^{+} e^{-}}$ reduces to:
\begin{equation} \label{eq:ratiopostelect2}
R_{e^{+} e^{-}}(\varepsilon,Q^2) = \frac{1-\delta_{2\gamma}}{1+\delta_{2\gamma}} \approx 1-2\delta_{2\gamma},
\end{equation} 
where $\delta_{2\gamma}$ is the fractional TPE correction for electron-proton scattering. Until recently, there was
only limited evidence for any non-zero TPE contribution from such comparisons~\cite{arrington04b}, as data
were limited to low $Q^2$ or large $\varepsilon$, where the TPE contributions appear to be small.  In 
addition, the details of the radiative corrections applied to these earlier measurements are not always
available, and it is not clear if the charge-even corrections were applied in all cases.  New
measurements~\cite{adikaram14, rachek14} have found more significant indications of TPE contributions at low
$\varepsilon$ and moderate $Q^2$, which are consistent with hadronic TPE calculations~\cite{blunden05a}.

\subsection{Extraction of the Form Factors and TPE}  \label{visit_interference}

To extract the proton form factors, we assume TPE contributions to the polarization data are negligible
and account for the contribution to $\sigma_{R}$ by adding the real function $F(\varepsilon,Q^2)$ to the Born
reduced cross section:
\begin{equation} \label{eq:reduced3}
\sigma_{R} = \big(G_M^p\big)^2\big[1 + \frac{\varepsilon}{\tau} \frac{R^2}{\mu_p^2}\big] + F(\varepsilon,Q^2),
\end{equation}
where $R=\mu_pG_E^p/G_M^p$ is the recoil polarization ratio.

We recently extracted the proton form factors and the TPE correction $a(Q^2)$ for large $Q^2$
values~\cite{qattan11a,qattan12} using the world data on $\sigma_R$ from
Refs.~\cite{janssens65,andivahis94,bartel73,litt70,berger71,qattan05,walker94,christy04}. The form
factors were extracted based on the parametrization from Borisyuk and Kobushkin (BK
parametrization)~\cite{borisyuk07}, where $\sigma_R$ is expressed in the following form
\begin{equation} \label{eq:Kobushkin3}
\sigma_{R} = \big(G_M^p\big)^2\big[ 1 + \frac{\varepsilon}{\tau} \frac{R^2}{\mu_p^2}\big] + 2 a(Q^2)(1-\varepsilon)\big(G_M^p\big)^2.
\end{equation} 
This form accounts for the experimental and theoretical constraints presented in Sec.~\ref{sec:TPE}. The
value for the ratio $R=\mu_p G_E^p/G_M^p$ is taken from a parameterization of the recoil polarization
data, for which we assume there is not TPE contribution. In the analysis of Refs.~\cite{qattan11a,
qattan12}, the linear parameterization of $R$ and its uncertainties from Ref.~\cite{gayou02} was used. 
An additional uncertainty was applied in the analysis of Ref.~\cite{qattan12} to provide more realistic
uncertainties at lower $Q^2$, but this was not included in Ref.~\cite{qattan11a}.

In this work, we extend the analysis of Ref.~\cite{qattan12} to lower $Q^2$ values by including
data from new, high-precision cross section measurements, and by improving the parameterization of $R$
and its uncertainty from polarization measurements at both low and high $Q^2$ values. Taking the
recoil-polarization and polarized-target results from Refs.~\cite{gayou01, puckett12, punjabi05,
strauch03, puckett10, ron11, jones06, maclachlan06, paolone10, zhan11, crawford07, pospischil01,
milbrath98}, we find that the form factor ratio is well parameterized as:
\begin{equation} \label{eq:JRA_RptNew}
R = \frac{1}{1+0.1430Q^2-0.0086Q^4+0.0072Q^6},
\end{equation} 
with  an absolute uncertainty in the fit given by: $\delta^2_R(Q^2) = (0.006)^2 +
(0.015\mbox{ln}(1+Q^2))^2$, with $Q^2$ in (GeV/c)$^2$.

\begin{figure}[!htbp]
\begin{center}
\includegraphics*[width=8.2cm]{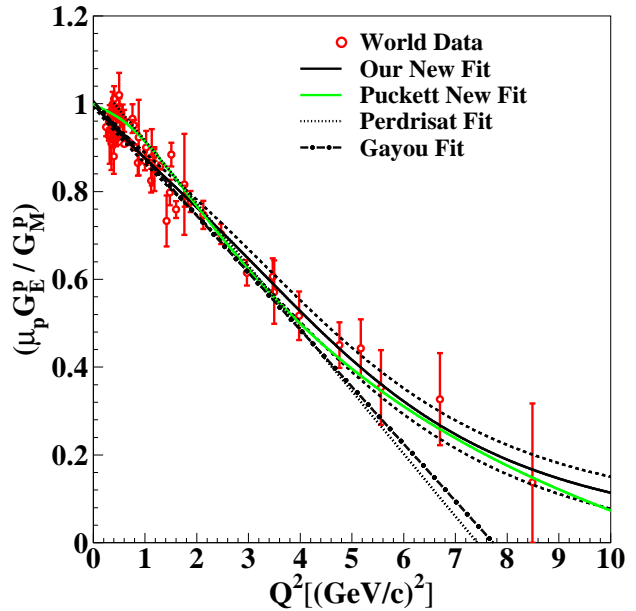}\\
\end{center}
\vspace{-0.5cm}
\caption{(Color online) The world data on the recoil-polarization ratio $\mu_{p}G_E^p/G_M^p$ from 
Refs.~\cite{gayou01, puckett12, punjabi05, strauch03, puckett10, ron11, jones06, maclachlan06, paolone10,
zhan11, crawford07, pospischil01, milbrath98} along with our new fit (black solid line) and its
uncertainty bands (black dashed lines) and previous fits~\cite{puckett12, perdrisat07, gayou02}.}
\label{fig:JRA_Rpt}
\end{figure}

An argument could be made that the uncertainty might be below 0.6\% for the low $Q^2$ values, especially
as we approach $Q^2=0$ where the ratio is known.  However, TPE calculations on the recoil-polarization
ratio show effects that can be up to ~0.5\%~\cite{blunden05a} at low $Q^2$. Because we neglect TPE
corrections to the polarization measurements, the uncertainties should be large enough to account for this.
Figure~\ref{fig:JRA_Rpt} shows our new global fit to the world data on the recoil-polarization ratio
$\mu_{p}G_E^p/G_M^p$ along with the fit's uncertainty bands.

\section{Results}

\subsection{Form Factors and TPE Contributions}

We extract the form factors and the TPE contributions using Eq.~(\ref{eq:Kobushkin3}), following the procedure of
Ref.~\cite{qattan12}. We obtain the value for $R$ and its uncertainty from the new parameterization presented
in Eq.~(\ref{eq:JRA_RptNew}). The cross section data included are those used in Ref.~\cite{qattan12} along with
the addition of low-$Q^2$ data from Refs.~\cite{bernauer10, janssens65}.

\begin{figure}[!htbp]
\begin{center}
\includegraphics*[width=8.2cm]{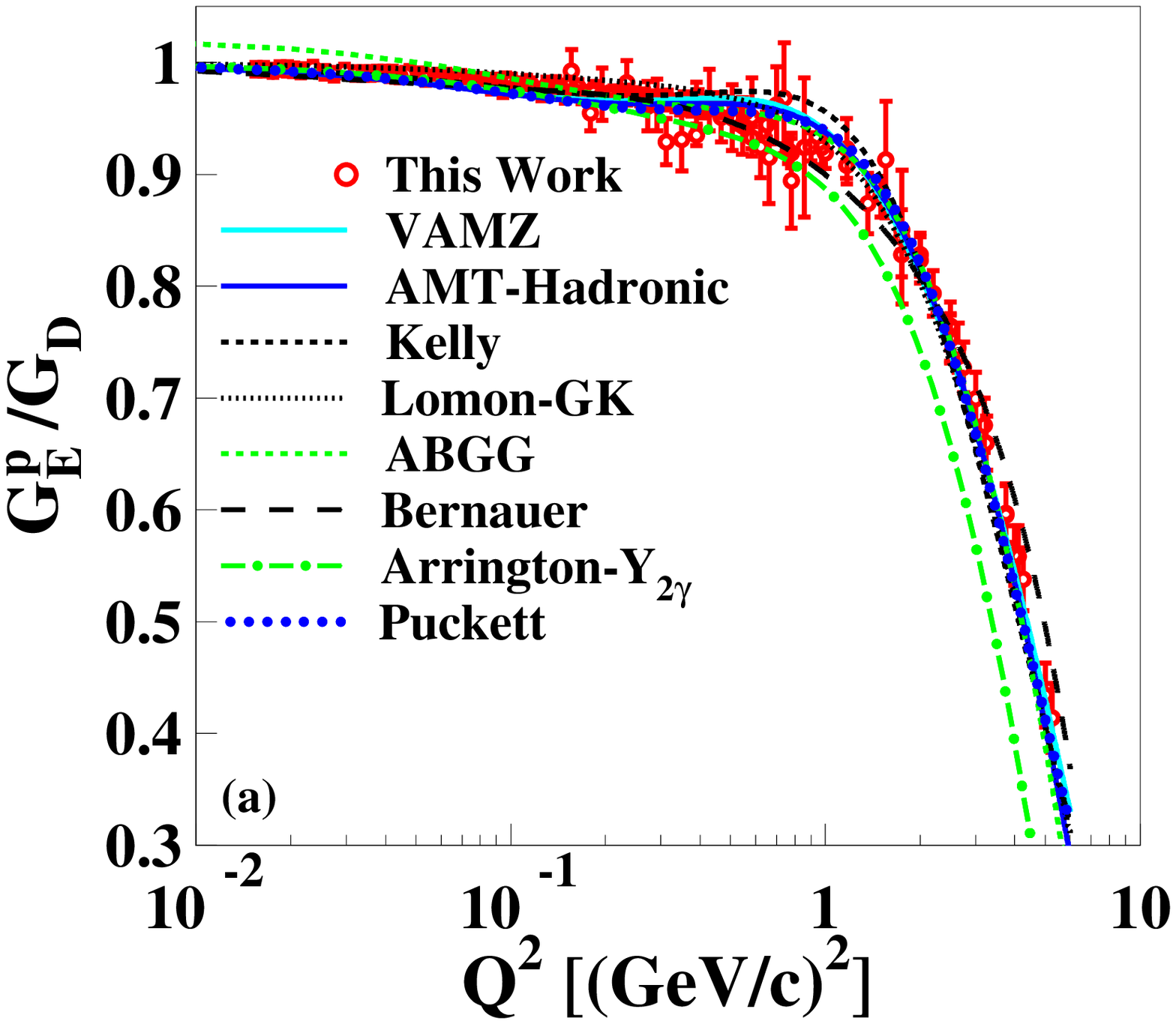}\\
\includegraphics*[width=8.2cm]{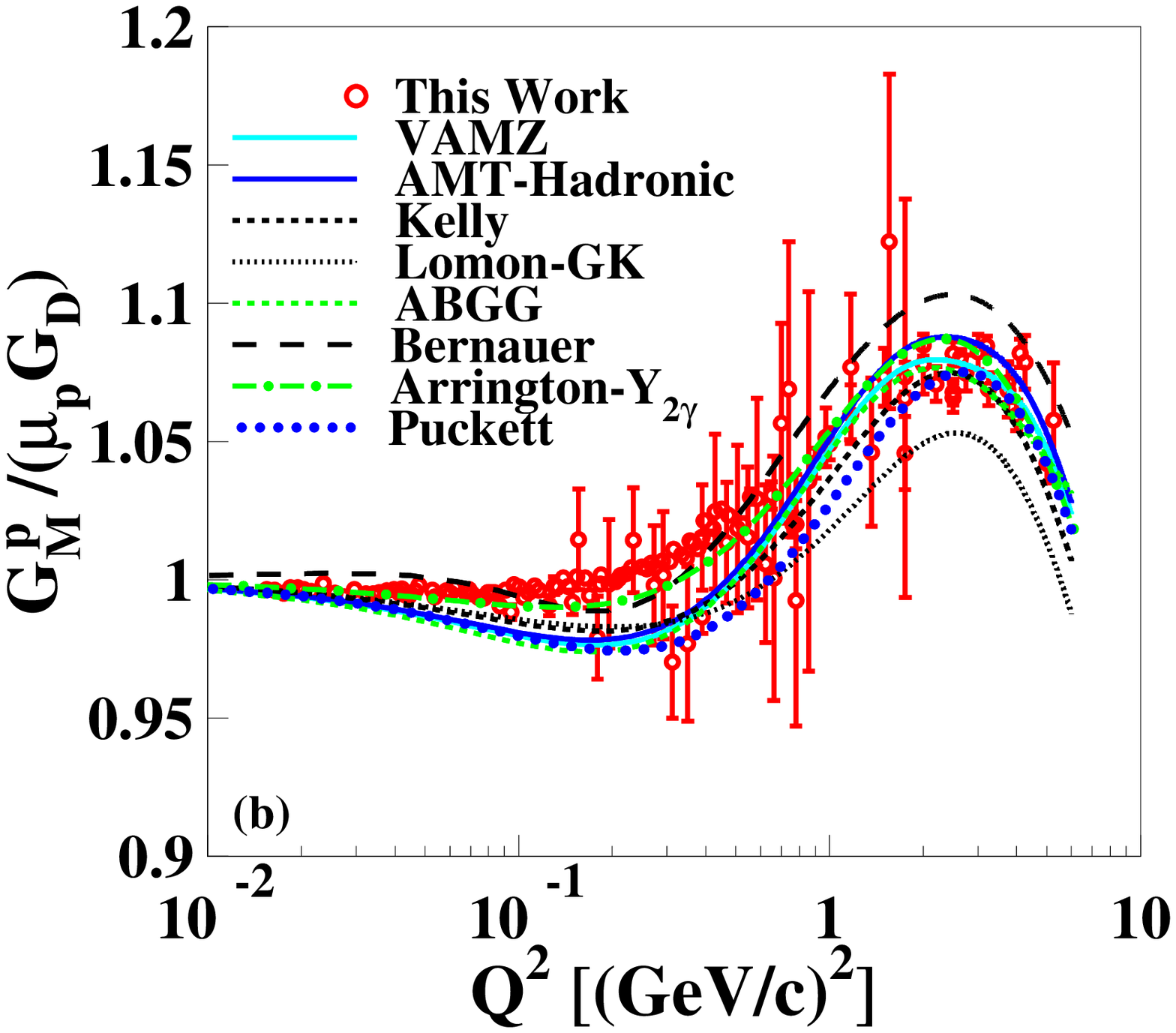}
\end{center}
\vspace{-0.5cm}
\caption{(Color online) $G_E^p/G_D$ (top) and $G_M^p/(\mu_{p}G_D)$ (bottom) as obtained using the BK 
parametrization from the data of Refs.~\cite{andivahis94,bartel73,litt70,berger71,qattan05,walker94,
christy04,janssens65,bernauer10}. In addition, we compare our results to extractions from several previous
TPE calculations and phenomenological fits: AMT~\cite{arrington07}, VAMZ~\cite{venkat11},
Kelly~\cite{kelly04}, Lomon-GK~\cite{lomon01}, ABGG~\cite{alberico09a}, Bernauer~\cite{bernauer14},
Arrington $Y_{2\gamma}$~\cite{arrington05}, and Puckett (the fit labeled ``new'' in Ref.~\cite{puckett12}).}
\label{fig:FFs}
\end{figure}

Figure~\ref{fig:FFs} shows the extracted proton form factors from this work; these form factors and the TPE
contribution are included in the online Supplemental Material~\cite{suppl2015}. We also show the form factors as
extracted after applying hadronic TPE calculations, labeled ``AMT-Hadronic''~\cite{arrington07} and ``VAMZ''
Ref.~\cite{venkat11}. In addition, we show fits to the form
factors from previous phenomenological analyses: ``Kelly''~\cite{kelly04}, ``Lomon-GK''~\cite{lomon01},
``ABGG''~\cite{alberico09a}, ``Bernauer''~\cite{bernauer14}, ``Arrington-$Y_{2\gamma}$''~\cite{arrington05},
and ``Puckett''~\cite{puckett12}.

While extractions using calculated TPE corrections and phenomenological-based fits are in reasonably good
agreement, extractions which combine cross section and polarization results but do not allow for an explicit
TPE correction, in particular the Kelly and Lomon extractions, tend to have larger differences.
The ``Arrington-$Y_{2\gamma}$'' fit is very different at large $Q^2$ because it is the only analysis that allows
for TPE contributions to the recoil polarization data, though the extraction of these terms is extremely
model dependent.

At low $Q^2$, our extraction of $G_M^p$ is significantly above most previous fits. This reflects the discrepancy
between the recent Mainz data which yields values of $G_M^p$ which are systematically 2--5\% larger than
previous world's data~\cite{bernauer14}. At low $Q^2$, this corresponds to only a small difference in the
cross section at large scattering angle, but for the larger $Q^2$ values of the Mainz experiment, this 
corresponds to a significant difference in the measured cross sections. Note that except for the Bernauer
result, most of the previous phenomenological extractions of the form factors and TPE contributions were
focused on large $Q^2$ values, and so did not always worry about how well the parameterizations of $R$
reproduced low $Q^2$ data.

\begin{figure}[!htbp]
\begin{center}
\includegraphics*[width=8.5cm]{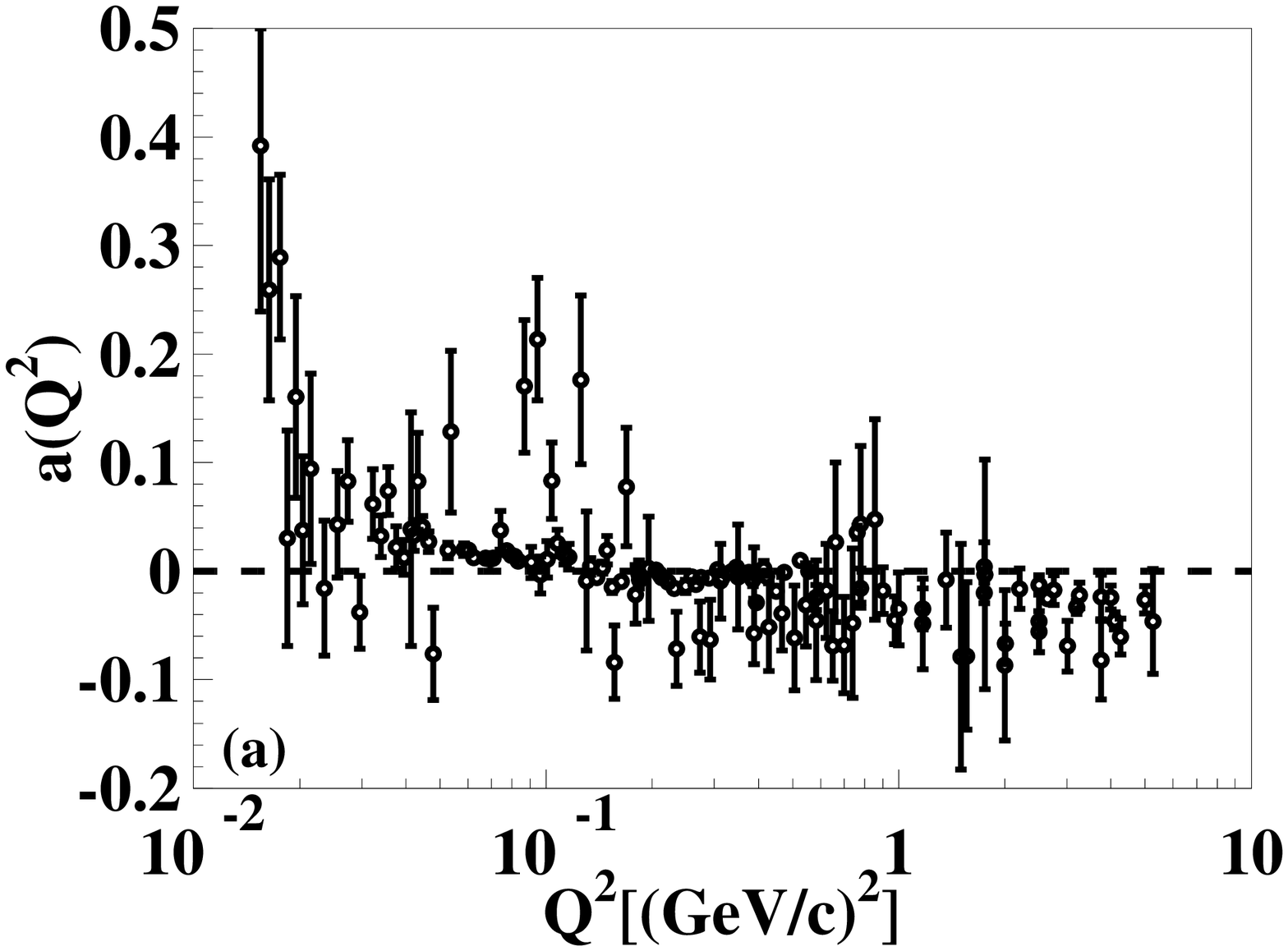}\\
\includegraphics*[width=8.5cm]{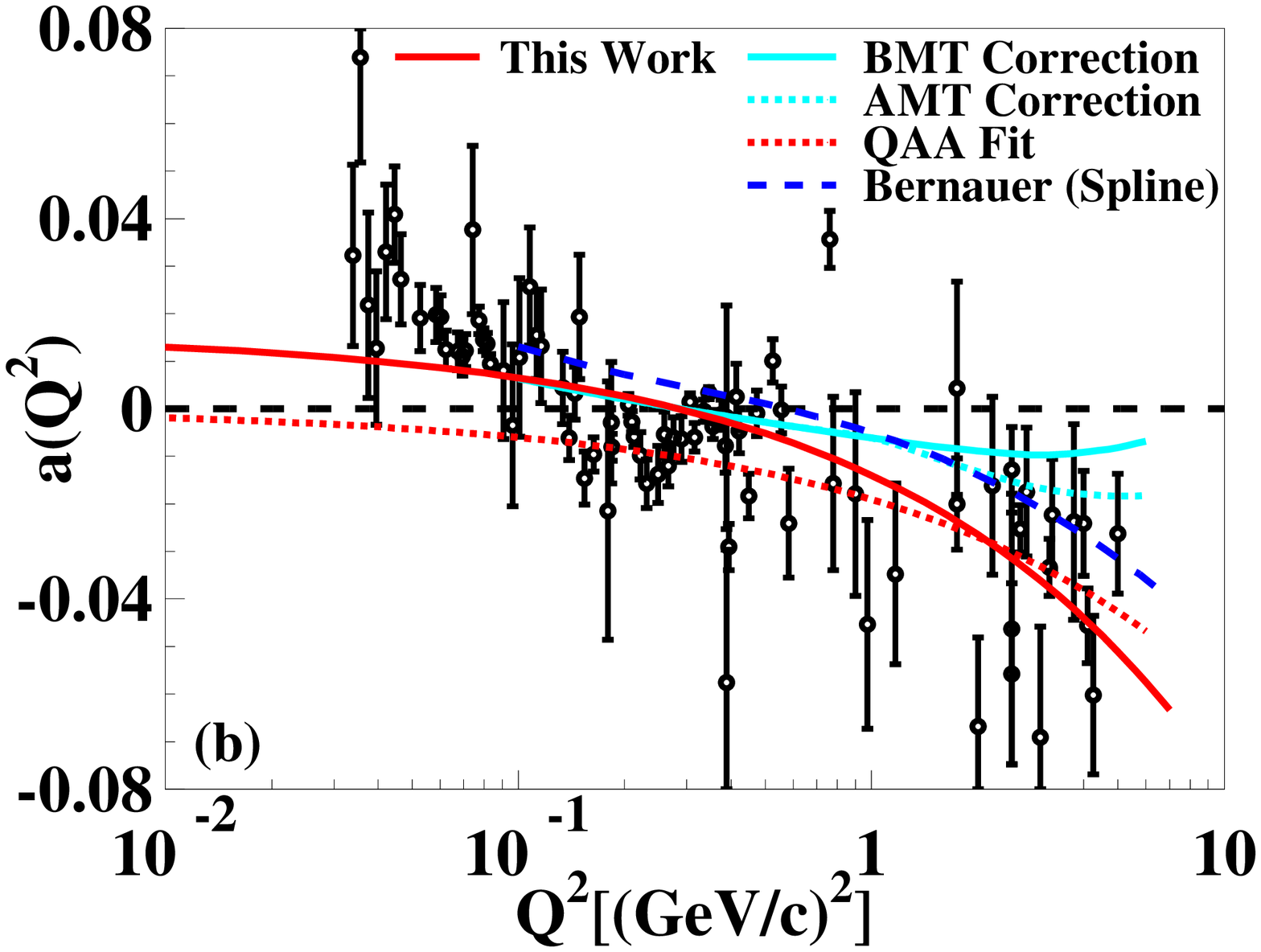}
\end{center}
\vspace{-0.5cm}
\caption{(Color online) The TPE term $a(Q^2)$ as obtained using the BK parametrization from the data of 
Refs.~\cite{andivahis94,bartel73,litt70,berger71,qattan05,walker94,christy04,janssens65,bernauer10}.
Also shown are a simple parameterization of our results and curves representing $a(Q^2)$ as determined in
previous analyses~\cite{blunden05a, arrington07, qattan11a, bernauer14}.  The bottom plot is on a smaller
vertical scale, excluding points with uncertainties above $\delta a=0.03$ for clarity, although they are
included in our fit. Note that we do not show extractions of $a(Q^2)$ for the calculations or
extraction of Ref.~\cite{bernauer14} at very low $Q^2$, as the $\varepsilon$ dependence is quite different
in our parameterization when the cross sections is dominated by the charge form factor (see text). }
\label{fig:aQ2fits}
\end{figure}

Figure~\ref{fig:aQ2fits} shows the TPE term $a(Q^2)$ extracted from this work. We also show parameterizations
of $a(Q^2)$ from the TPE hadronic calculations of Ref.~\cite{blunden05a} (BMT) and
Ref.~\cite{arrington07} (AMT), which adds and additional phenomenological TPE contribution at higher $Q^2$,
and from previous phenomenological extractions~\cite{qattan11a,bernauer14}. At high-$Q^2$ values, the TPE
term is consistent with our previous extraction and shows an increase in magnitude with increasing
$Q^2$. The previous extraction~\cite{qattan11a} was not well constrained at low $Q^2$, with cross section data limited to
$Q^2 \geq 0.39$ and a parameterization~\cite{gayou02} of the polarization data that was not well constrained
at low $Q^2$. So below $Q^2 \approx 1$~(GeV/c)$^2$, the behavior was driven by the fitting function which
took $a(Q^2)$ proportional to $\sqrt{Q^2}$. In the present work, the inclusion of the Mainz data and low
$Q^2$ polarization provide meaningful uncertainties for $a(Q^2)$ below $Q^2=1$~(GeV/c)$^2$ and show a change
in sign at low $Q^2$, as seen in previous low-$Q^2$ TPE calculations~\cite{blunden05a, borisyuk07, arrington04c,
arrington13}.

While the Mainz data show indications of structure in $a(Q^2)$, the uncertainties for these data are underestimates
of the true uncertainties. The quoted uncertainties on the individual cross sections do not include correlated
systematic effects, which are a significant contribution to the total uncertainty in their final form factor
parameterization. We fit to the $Q^2$ dependence of the extracted TPE contributions, adding an additional
uncertainty, $da=0.01$, to each point to help offset the impact of the underestimated uncertainties in our extraction
from the Mainz data. We obtain $a(Q^2) = 0.016 - 0.030 \sqrt{Q^2}$, with $Q^2$ in (GeV/c)$^2$. Note that for
our parameterization of TPE contributions, $F(Q^2,\varepsilon)/(G_M^p)^2$ is a linear function in $\varepsilon$,
while the $Q^2 \to 0$ limit~\cite{mckinley48, arrington11b} is more nearly linear when taken as a ratio to the
reduced cross section, $(G_M^p)^2+(\varepsilon/\tau)(G_E^p)^2$, and so the behavior at small $Q^2$, where $1/\tau$
becomes large, is very different. This can be seen more clearly in examining the $\varepsilon$ dependence of the
TPE contributions at very low $Q^2$, as shown in Fig.~\ref{fig:LowQ2Rpm}.

\subsection{Ratio of Positron-Proton to Electron-Proton Elastic Scattering Cross Sections $R_{e^{+} e^{-}}$} \label{Rpm_Ratio}

\begin{figure}[!htbp]
\begin{center}
\begin{tabular}{c c}
\includegraphics*[width=4.1cm]{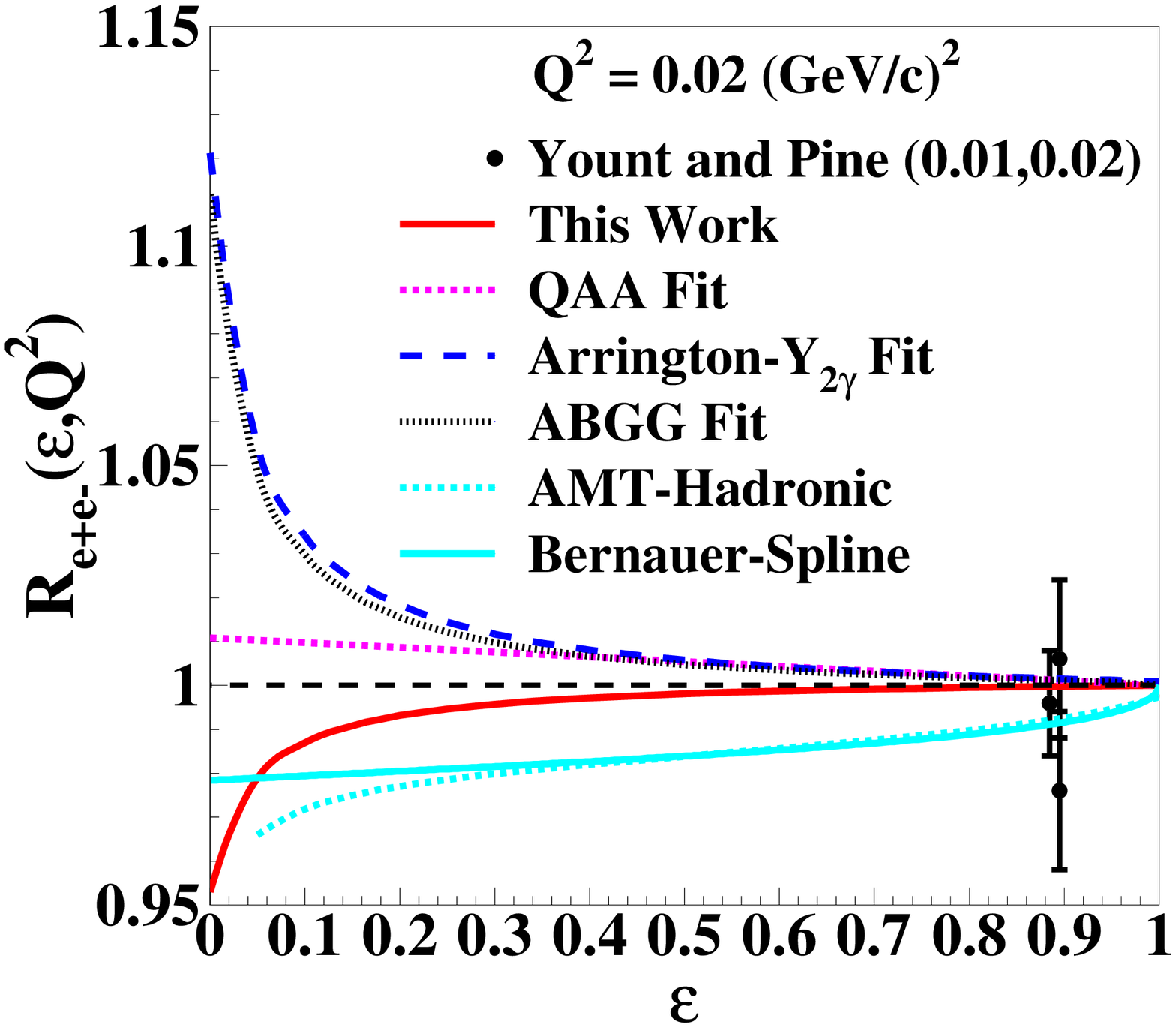} &
\includegraphics*[width=4.1cm]{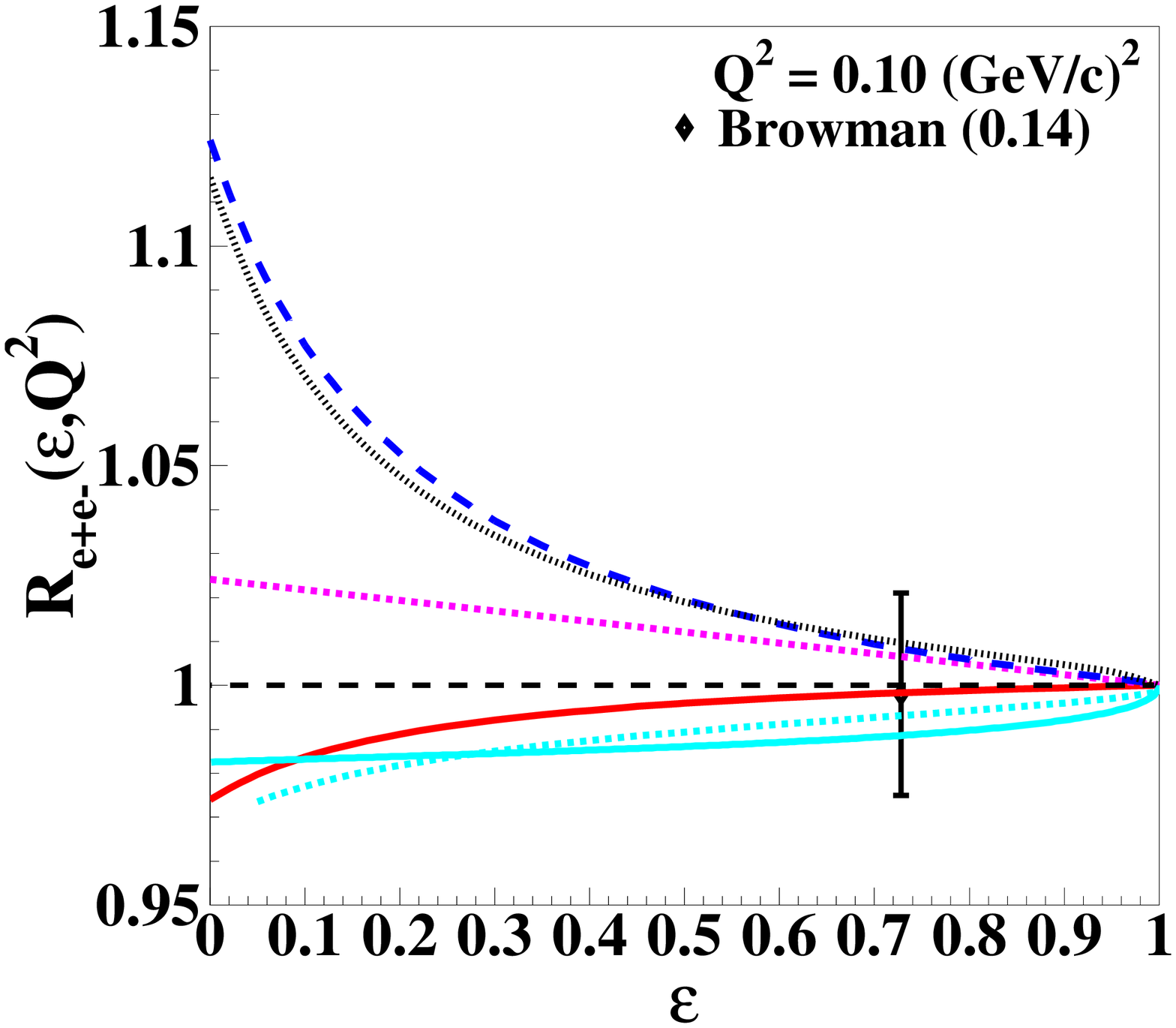} \\
\includegraphics*[width=4.1cm]{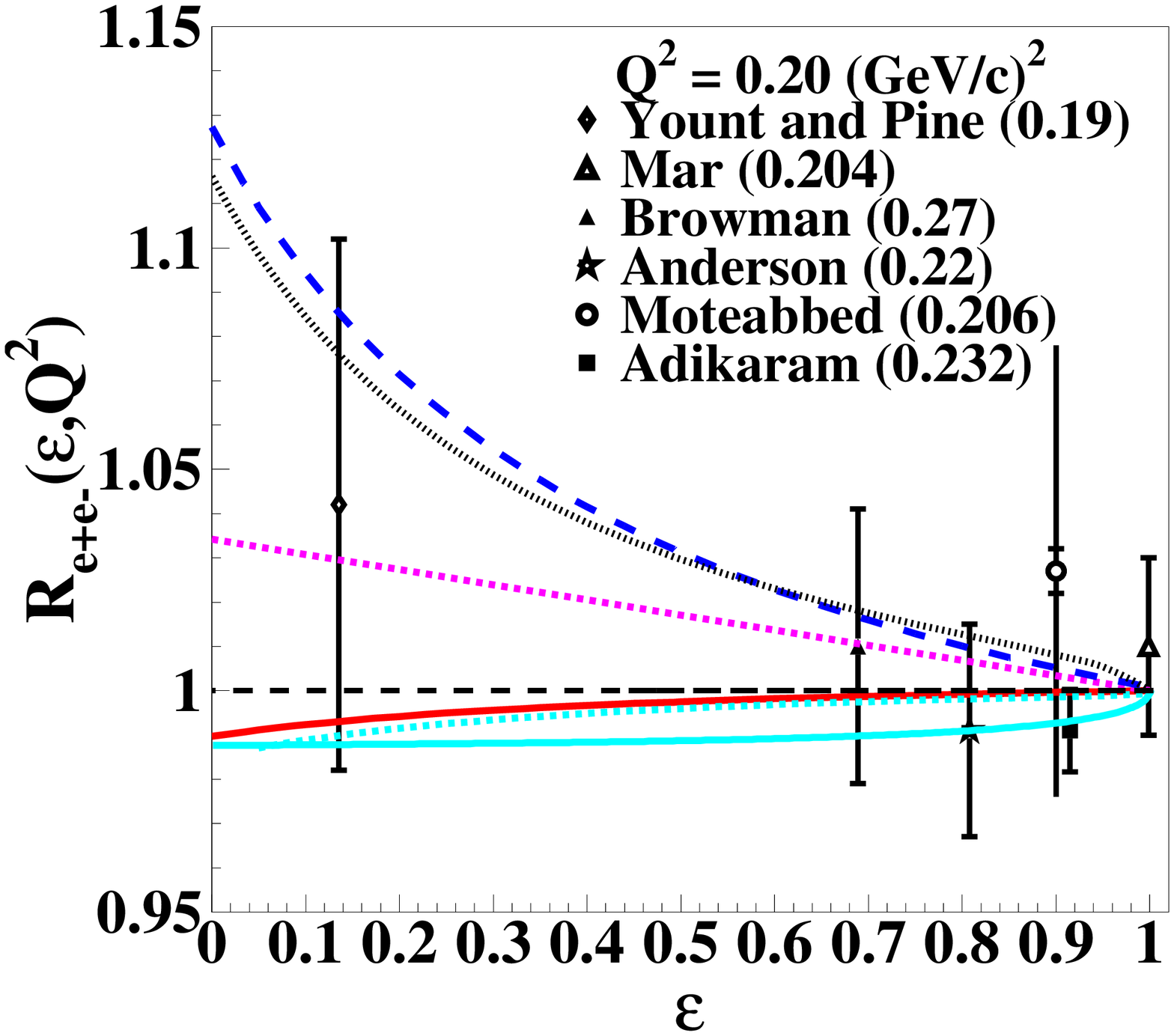} &
\includegraphics*[width=4.1cm]{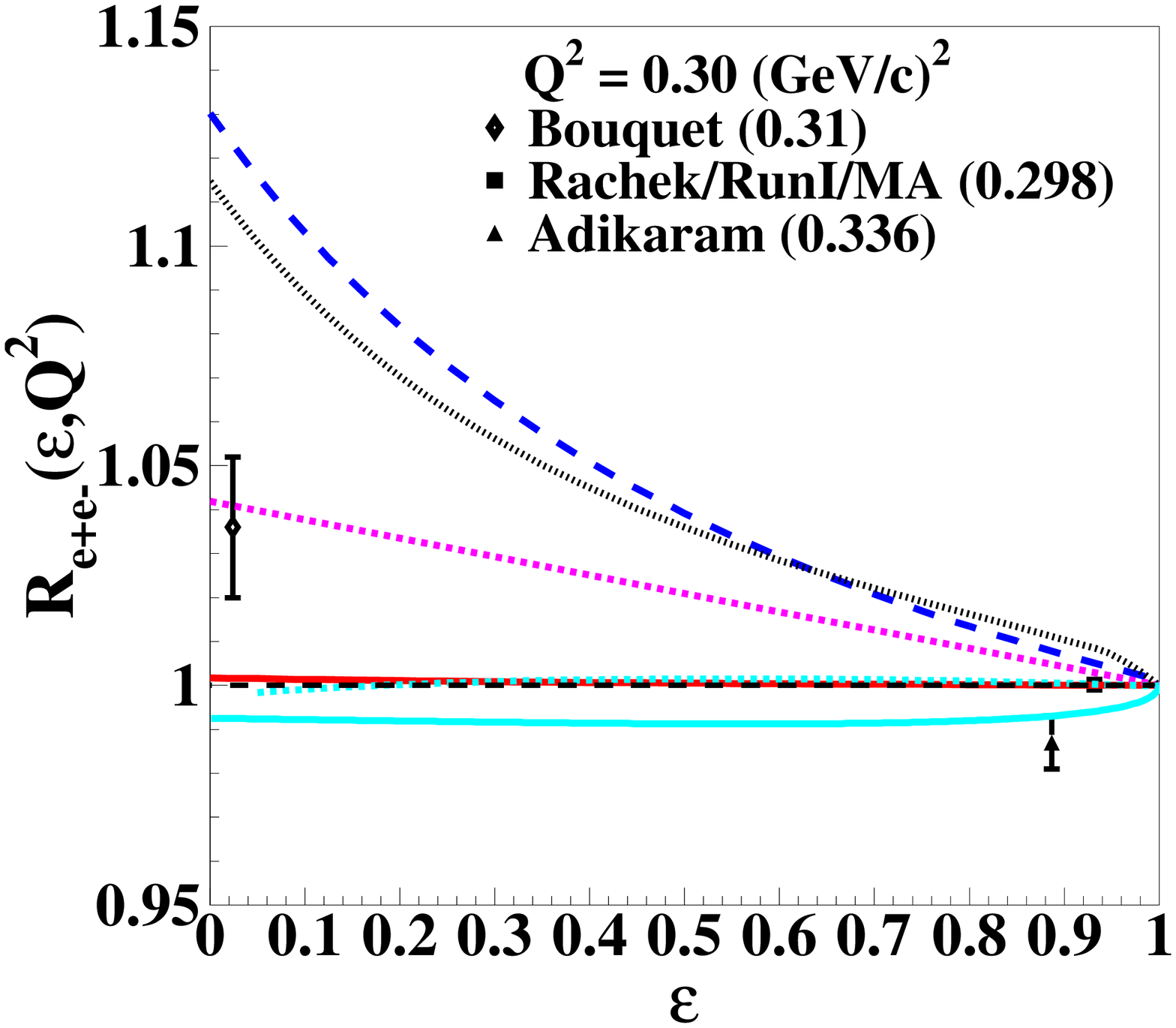} \\
\includegraphics*[width=4.1cm]{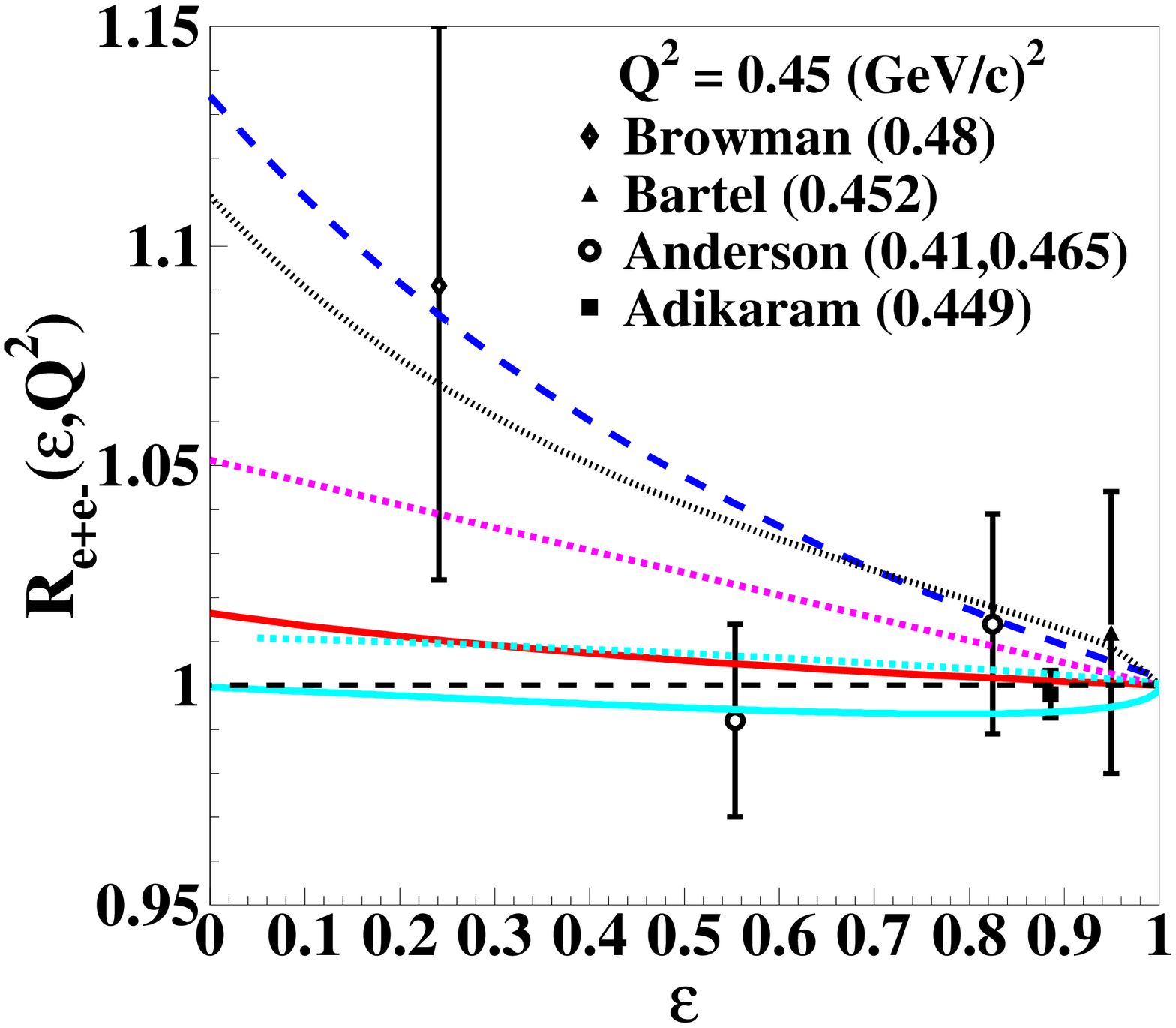} &
\includegraphics*[width=4.1cm]{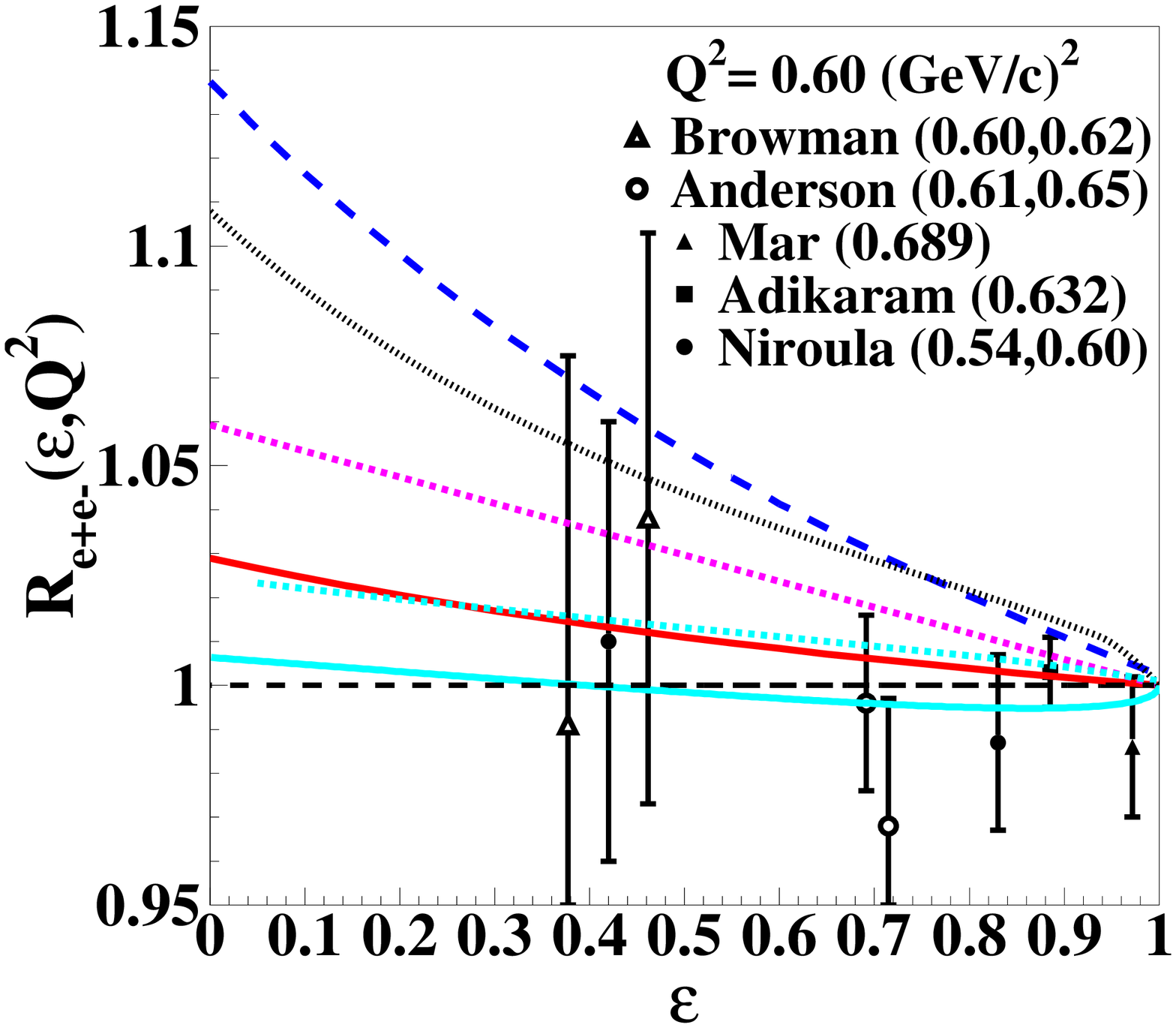} \\
\includegraphics*[width=4.1cm]{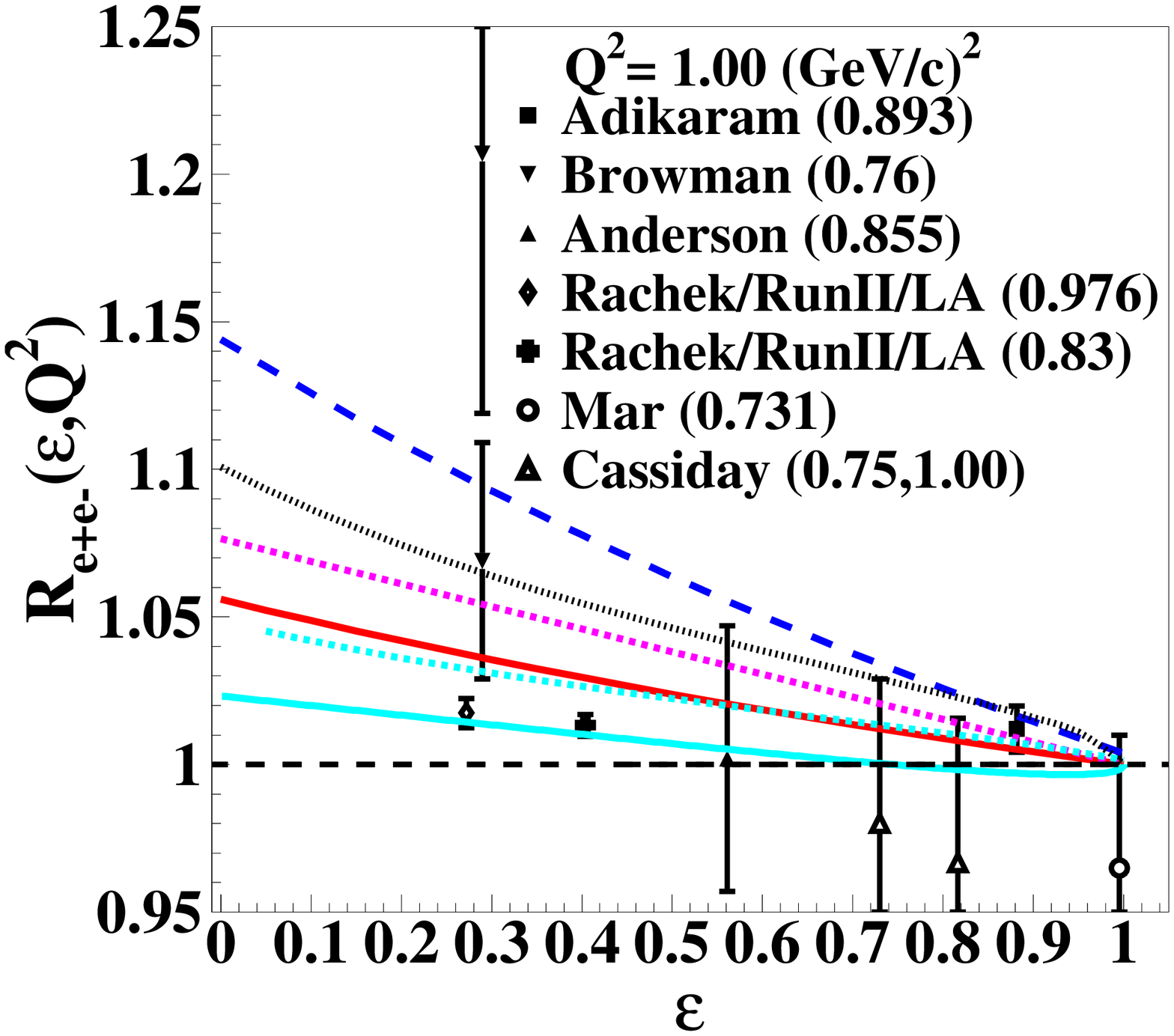} &
\includegraphics*[width=4.1cm]{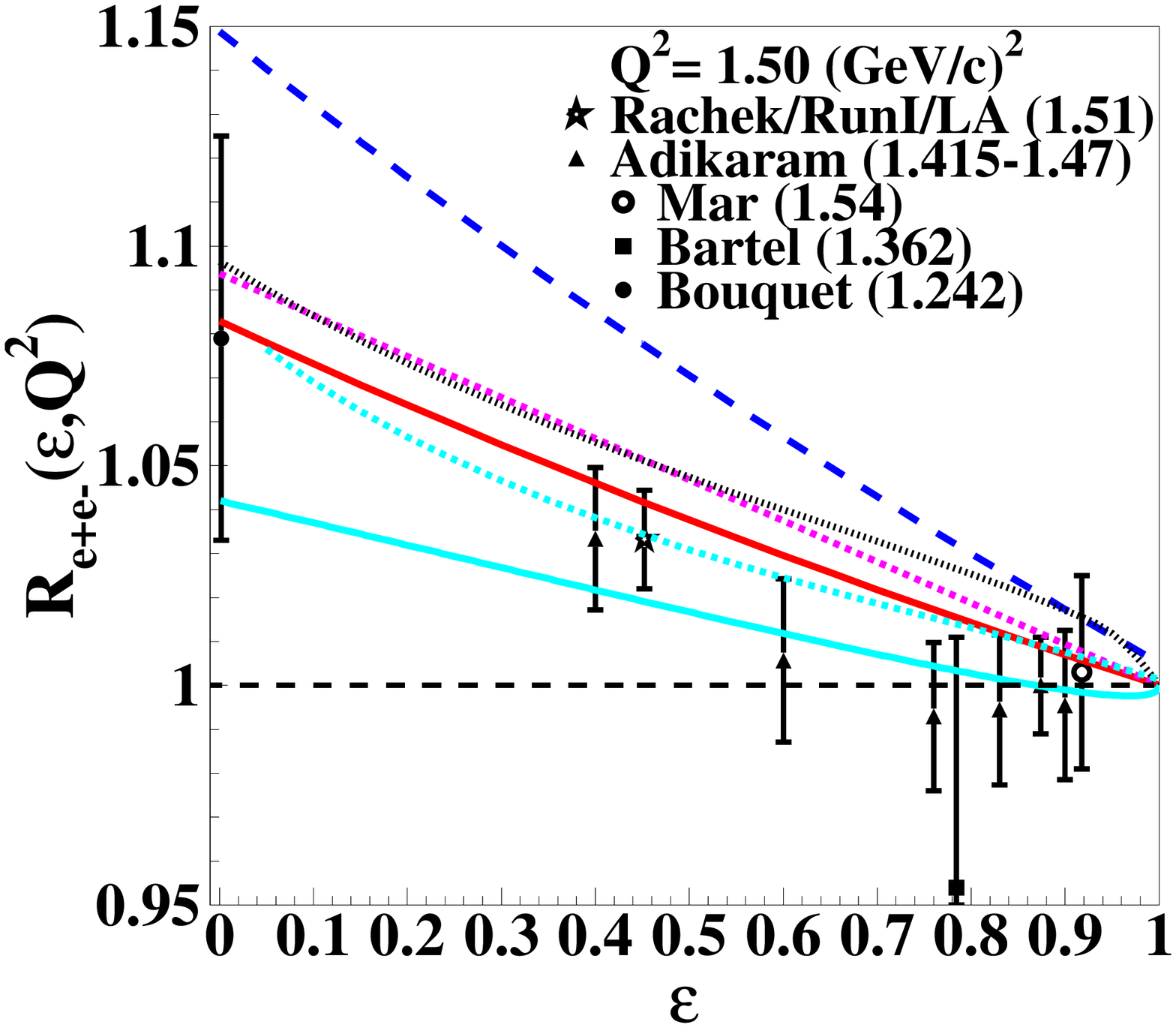} \\
\vspace{-0.5cm}
\end{tabular}
\end{center}
\caption{(Color online) The ratio $R_{e^{+} e^{-}}(\varepsilon,Q^2)$ as determined from our parameterization
of $a(Q^2)$ at the $Q^2$ value listed in the figure. The other curves show previous
calculations~\cite{arrington07} or phenomenological extractions~\cite{qattan11a, arrington05, alberico09a, bernauer14}.
The data points are direct measurements of $R_{e^{+} e^{-}}$~\cite{adikaram14, rachek14, yount62, browman65, mar68, anderson66,
anderson68, bartel67, bouquet68, moteabbed13, niroulaphd10, cassiday67}. For the
world data, the measurement and $Q^2$ value(s) in (GeV/c)$^2$ are given.}
\label{fig:LowQ2Rpm}
\end{figure}

From the TPE contributions extracted based on the parameterization of Eq.~(\ref{eq:Kobushkin3}), the corrected
ratio of positron-to-electron scattering cross sections is
\begin{equation} \label{eq:ratiopostelect3}
R_{e^{+} e^{-}}(\varepsilon,Q^2) = \frac{1-\delta_{2\gamma}}{1+\delta_{2\gamma}} \approx 1-\frac{4a(Q^2)(1-\varepsilon)}{(1+\frac{\varepsilon R^2}{\tau \mu_p^2})}.
\end{equation}
Figure~\ref{fig:LowQ2Rpm} shows the ratio $R_{e^{+} e^{-}}$ as a function of $\varepsilon$ extracted for a range of
$Q^2$ values.  The solid red curve represents the ratio from our new parameterization of $a(Q^2)$, and the others
show the results of TPE hadronic calculations from Refs.~\cite{arrington07,blunden05a} and several phenomenological
fits from Refs.~\cite{arrington05,alberico09a,qattan11a,bernauer14}.  The data points are world data on the ratio
$R_{e^{+} e^{-}}$ from Refs.~\cite{adikaram14, rachek14, yount62, browman65, mar68, anderson66, anderson68,
bartel67, bouquet68, moteabbed13, niroulaphd10, cassiday67}. Two recent measurements~\cite{adikaram14, rachek14}
have provided significantly more precise measurements at $Q^2 \approx 1.0$ and 1.5~(GeV/c)$^2$, providing evidence
for non-zero TPE at larger $Q^2$ values and a change of sign from the exact calculation at $Q^2=0$, consistent with
what we observe.  Our results are slightly larger than the direct measurements at 1 (GeV/c)$^2$ from
Ref.~\cite{rachek14}, but otherwise in very good quantitative agreement with existing data.

Note that parameterizations where the TPE contribution is similar to that of Eq.~(\ref{eq:Kobushkin3}), i.e. a linear
function times $(G_M^p)^2$, the low $Q^2$ results for $R_{e^{+} e^{-}}$ will have a strong non-linear behavior. The
TPE contribution relative to $(G_M^p)^2$ is linear, but $(G_E^p)^2$ dominates the cross section at very low $Q^2$
except for $\varepsilon \to 0$, strongly suppressing TPE as a fractional contribution as one moves away from
$\varepsilon=0$. This could be fixed by modifying the functional form, e.g. using a linear function in $\varepsilon$
times the full reduced cross section~\cite{bernauer14}. However, because the results are similar everywhere except at
very low $Q^2$, where there is little data, we consider this parameterization sufficient for the present analysis.

\subsection{Flavor Separation of the Nucleon Form Factors} \label{FFsFlavor}

Examinations of the flavor-separated form factors of the nucleon focusing on high $Q^2$~\cite{cates11,qattan12} have
provided several interesting observations. Below we summarize some of the main conclusions from these analyses:\\

(1) The down-quark contributions to Dirac, $F_1$, and Pauli, $F_2$, form factors deviate from the expected
$1/Q^4$ scaling~\cite{cates11}, with small differences between the $Q^2$ dependence in $F_1$ and $F_2$ for both the
up- and down-quark contributions.

(2) The up- and down-quark yield very different contributions~\cite{qattan12} to $G_E/G_M$ and $F_2/F_1$. The
strong linear falloff with $Q^2$ in the ratio $G_E^p/G_M^p$ is not seen in either the up- or down-quark
contributions, but mainly arises due to a cancellation between a weaker $Q^2$ dependence for the up-quark and a
negative but relatively $Q^2$-independent contribution from the down-quark. 

(3) The more recent analysis~\cite{qattan12} shows some differences from the original work by
Cates~\etal~\cite{cates11}, referred to as ``CJRW'' throughout the text. The difference are associated with
the approximations made in the CJRW analysis, which neglected TPE effects and included only the uncertainty
associated with the neutron charge form factor.  The treatment of TPE contributions yields small but clear
differences, mainly at lower $Q^2$ values, up to $\approx 1.5$~(GeV/c)$^2$, while the addition of uncertainties
associated with TPE and all of the form factors yields somewhat larger uncertainties in most form factors, and
provides an estimate for the uncertainties in the magnetic form factors absent in the CJRW analysis. 

\begin{figure}[!htbp]
\begin{center}
\includegraphics*[width=8.3cm, height=7cm]{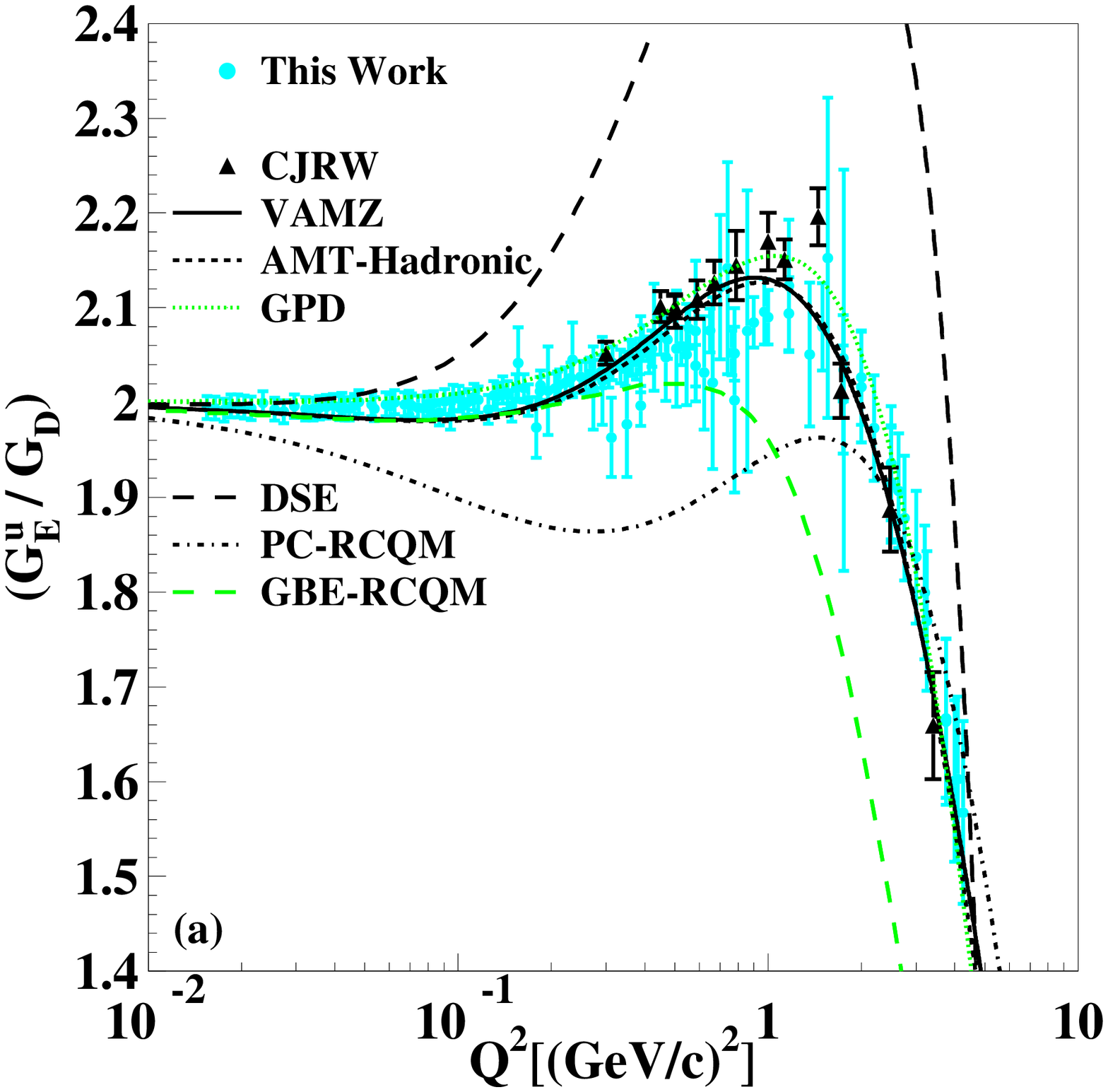}\\
\includegraphics*[width=8.3cm, height=7cm]{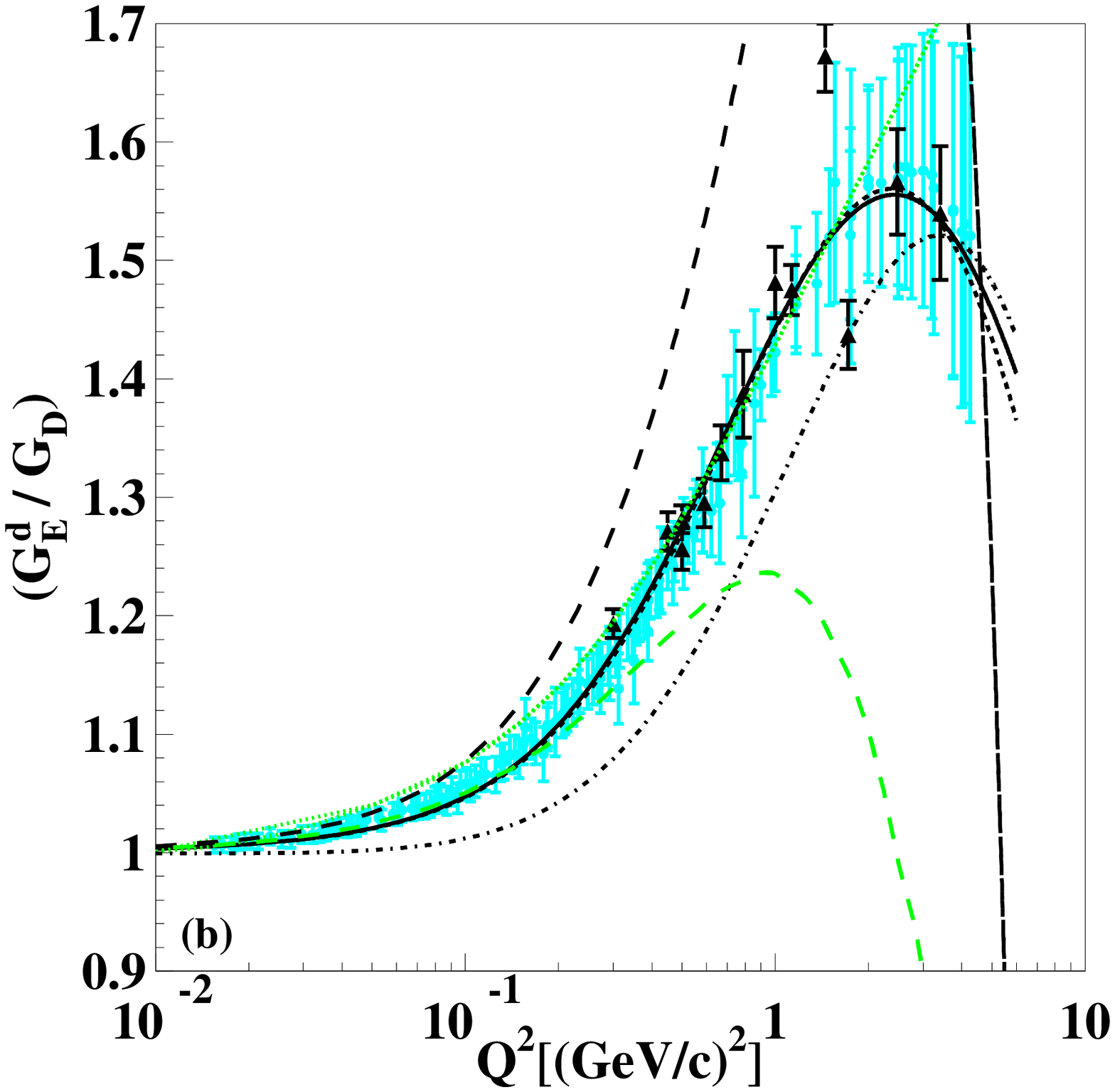}
\end{center}
\vspace{-0.5cm}
\caption{(Color online) ($G_E^u/G_D$) (top) and ($G_E^d/G_D$) (bottom) as a function of $Q^2$ 
from the data of Refs.~\cite{andivahis94,bartel73,litt70,berger71,qattan05,walker94,christy04,janssens65,bernauer10}.
Also shown the CJRW extractions~\cite{cates11} (open triangles), the AMT-Hadronic~\cite{arrington07} and 
VAMZ~\cite{venkat11} fits, and the values from the GPD~\cite{gonzalez13}, DSE~\cite{cloet09}, PC-RCQM~\cite{cloet12}, 
and GBE-RCQM~\cite{rohrmoser11a, rohrmoser13} models. Note that for the proton, the up- and down-quark
contributions have weighting factors of 2/3 and -1/3 (Eq. (\ref{eq:FFs_Flavor1})).}
\label{fig:GEp_ud}
\end{figure}

\begin{figure}[!htbp]
\begin{center}
\includegraphics*[width=8.3cm, height=7cm]{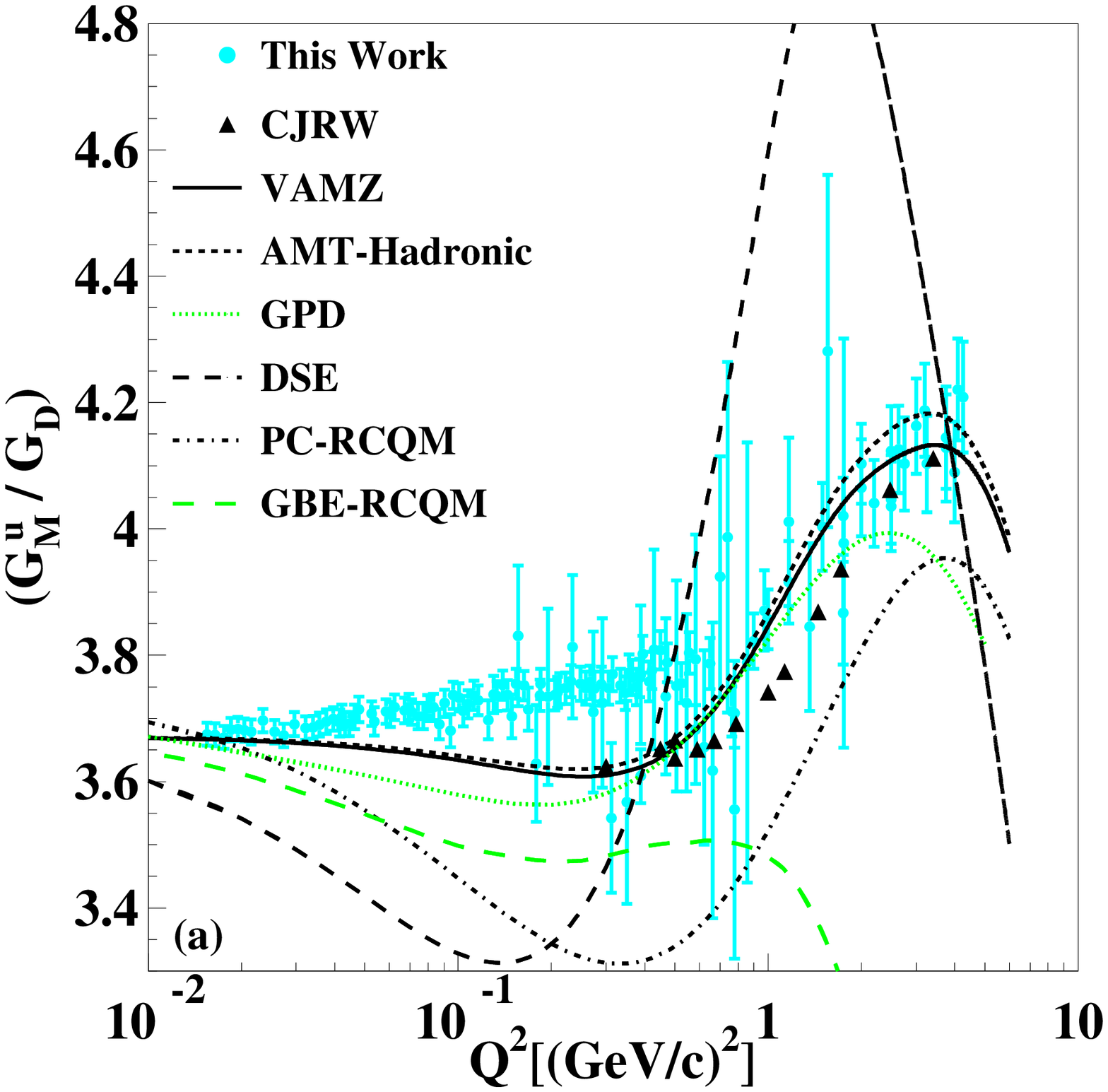}\\
\includegraphics*[width=8.3cm, height=7cm]{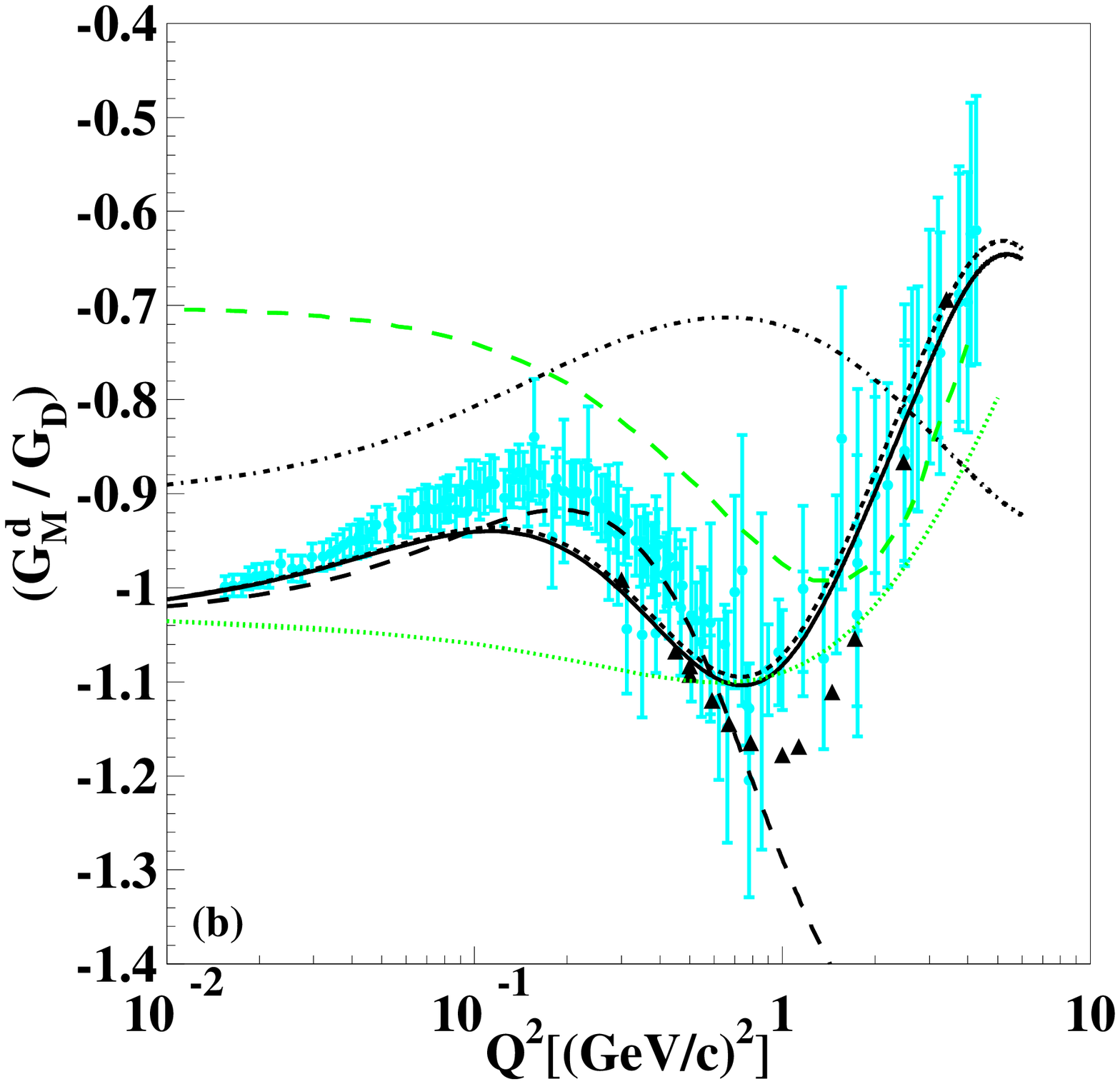}
\end{center}
\vspace{-0.5cm}
\caption{(Color online) ($G_M^u/G_D$) (top) and ($G_M^d/G_D$) (bottom) as a function of $Q^2$ 
from the data of Refs.~\cite{andivahis94,bartel73,litt70,berger71,qattan05,walker94,christy04,janssens65,bernauer10}.
Also shown the CJRW extractions~\cite{cates11} (open triangles), the AMT-Hadronic~\cite{arrington07} and 
VAMZ~\cite{venkat11} fits, and the values from the GPD~\cite{gonzalez13}, DSE~\cite{cloet09}, PC-RCQM~\cite{cloet12}, 
and GBE-RCQM~\cite{rohrmoser13} models. Note that for the proton, the up- and down-quark
contributions have weighting factors of 2/3 and -1/3 (Eq. (\ref{eq:FFs_Flavor1})).}
\label{fig:GMp_ud}
\end{figure}

Figures~\ref{fig:GEp_ud} and~\ref{fig:GMp_ud} show the flavor-separated form factors extracted from this work;
the extracted values are included in the online Supplemental Material~\cite{suppl2015}. The updated
parameterization of the polarization-transfer measurements of $\mu_pG_E^p/G_M^p$ yields a small difference in the
data above 5~(GeV/c)$^2$, but it is always a factor of 2-3 below the quoted uncertainties. Note that because the
CJRW analysis includes only the uncertainties from $G_E^n$, the flavor-separated extraction of $G_M$ is quoted
without uncertainty. In our analysis, uncertainties are included for all form factors, but we use a
parameterization for the uncertainty on $G_E^n$ and $G_M^n$, taken from Ref.~\cite{qattan12}. Because of this,
cases where the uncertainty is dominated by neutron data yield points with large error bars but small scatter
between points, as only the proton uncertainties are independent for each $Q^2$ values.

The results are compared to the original CJRW results and two recent extractions of the nucleon elastic form
factors which apply calculated TPE contributions~\cite{arrington07,venkat11}.  Also shown are calculations based on
dressed-quarks contribution within the frame work of Dyson-Schwinger equation ``DSE'' from Ref.~\cite{cloet09},
pion-cloud relativistic constituent quark model ``PC-RCQM'' from Ref.~\cite{cloet12}, relativistic constituent
quark model whose hyperfine interaction is derived from Goldstone-boson exchange ``GBE-RCQM'' from
Ref.~\cite{rohrmoser11a,rohrmoser13}, and generalized parton distributions with incorporated Regge
contribution that apply diquark models ``GPD'' from Ref.~\cite{gonzalez13}.

We start by examining the contributions to $G_E^p$. The down-quark contribution is much smaller than the up-quark
contribution at all $Q^2$ values. Up to $Q^2 \approx 1$~(GeV/c)$^2$, both $G_E^d/G_D$ and $G_E^u/G_D$ increase with
$Q^2$, deviating noticeably from the dipole form at $Q^2 \approx 0.10$~(GeV/c)$^2$ and $Q^2 \approx 0.40$~(GeV/c)$^2$,
respectively. At higher $Q^2$ values, $G_E^u/G_D$ decreases rapidly while $G_E^d/G_D$ continues to increase,
leading to the significant falloff in $G_E^p/G_D$ after applying the charge weighting of the up- and down-quark
contributions. While most of the calculations give a reasonable qualitative description of the data, all showing
a rise and then fall of both $G_E^d/G_D$ and $G_E^u/G_D$ with increasing $Q^2$ values, the AMT-Hadronic and VAMZ
parameterizations, as well as the GPD model, which also fits to the measured form factors, provide the best
description of the data.

This work allows for a more detailed examination of the low-$Q^2$ region, which is sensitive to the RMS radius of
the up- and down-quark distributions. At very low $Q^2$ values, $G_E^u$ is consistent with the dipole fit, while
$G_E^d$ falls less slowly than the dipole, yielding an increase in the ratio $G_E^d/G_D$. The proton RMS charge
radius as determined from electron scattering is $r_E \approx 0.88$~fm~\cite{zhan11, bernauer14}. The dipole form
corresponds to an RMS radius of 0.811~fm, suggesting an RMS radius of approximately 0.81~fm for the up-quark
distribution and below 0.81~fm for the down quarks. This gives $r_d < r_u < r_p$ for the charge radii, i.e. both the
up- and down-quark distributions are more localized than the overall proton charge distribution, due to cancellation
of the up- and down-quark contributions in the total proton charge distribution. Note that while the proton
charge radius as extracted from muonic hydrogen measurements~\cite{pohl10, antognini13} yield $r_E \approx
0.84$~fm, it is more natural to compare scattering results of the proton and flavor-separated contributions,
although the up- and down-quark radii are also smaller than muonic hydrogen charge radius.

For $G_M^p$, the up-quark contribution is again dominant, even more so than for the charge form factor.
Our results are in relatively good
agreement with the CJRW analysis for $Q^2$ above 2~(GeV/c)$^2$, though with a small offset associated with TPE.
For $Q^2 \ltorder 1$~(GeV/c)$^2$, we find noticeably larger values for both the up- and down-quark contributions
to the magnetic form factor. This difference comes from the Mainz data which yields values of $G_M^p$ which are
systematically larger than previous world's data, as seen in Fig.~\ref{fig:FFs}. Thus, it is not surprising that
there is an inconsistency between the extraction from these data and other measurements, in particular for the
up-quark contribution. For both the flavor-separated charge and magnetic form factors, the global
fits~\cite{arrington07,venkat11} describe the data (excluding the results from the Mainz measurement) well.
The models with many free parameters~\cite{cloet12, gonzalez13} tend to do a better job than the models with few
or no parameters adjusted to match the form factor data ~\cite{cloet09, rohrmoser13}, although the GPD
parameterization provides a significantly worse description of the magnetic form factor. 

Focusing on the very low $Q^2$ data, we see that $G_M^u$ falls more slowly than the dipole, indicating a
magnetic radius smaller than the 0.81~fm associated with the dipole form.  The magnitude of the down quark
contribution falls more rapidly with $Q^2$, indicating a larger radius, although because $G_M^d$ is negative
this corresponds to an increase in the ratio. The uncertainty on the proton's magnetic radius is significantly
larger than for the electron radius~\cite{zhan11, bernauer14, antognini13}, so while the magnetization distribution
is clearly larger for the down quarks than for the up quarks, opposite of what is observed for the charge radius,
it is difficult to determine how these compare to the overall proton magnetization radius.

\subsection{Flavor-dependent contributions to $\sigma_R$} \label{SigR_Flavor}
 
Using Eqs.~(\ref{eq:FFs_Flavor1}) and~(\ref{eq:reduced3}), the reduced cross section in the Born
approximation can be written as
\begin{equation} \label{eq:SigRFlavor1}
\sigma_R = \big(\frac{2}{3} G_M^u - \frac{1}{3} G_M^d\big)^2 +\frac{\varepsilon}{\tau}
\big(\frac{2}{3} G_E^u - \frac{1}{3} G_E^d\big)^2\nonumber 
\end{equation}
which can be separated into terms coming from just the up-quark or down-quark contributions and the
up-down interference term:
\begin{eqnarray} \label{eq:SigRFlavor3}
\sigma^{(u)}= \big(\frac{2}{3}G_M^u\big)^2 
+ \frac{\varepsilon}{\tau}\big(\frac{2}{3}G_E^u\big)^2, ~~~~~~~~~~~~~~~~~~~~~~\nonumber \\                       
\sigma^{(d)}= \big(\frac{-1}{3}G_M^d\big)^2 
+ \frac{\varepsilon}{\tau}\big(\frac{-1}{3}G_E^d\big)^2, ~~~~~~~~~~~~~~~~~~\nonumber \\
\sigma^{(u)\times(d)}= 2\big(\frac{2}{3}G_M^u\big)\big(\frac{-1}{3}G_M^d\big)
+\frac{2\varepsilon}{\tau}\big(\frac{2}{3}G_E^u\big)\big(\frac{-1}{3}G_E^d\big).\nonumber \\
\end{eqnarray}

\begin{figure}[!htbp]
\begin{center}
\begin{tabular}{c c}
\includegraphics*[width=4.3cm]{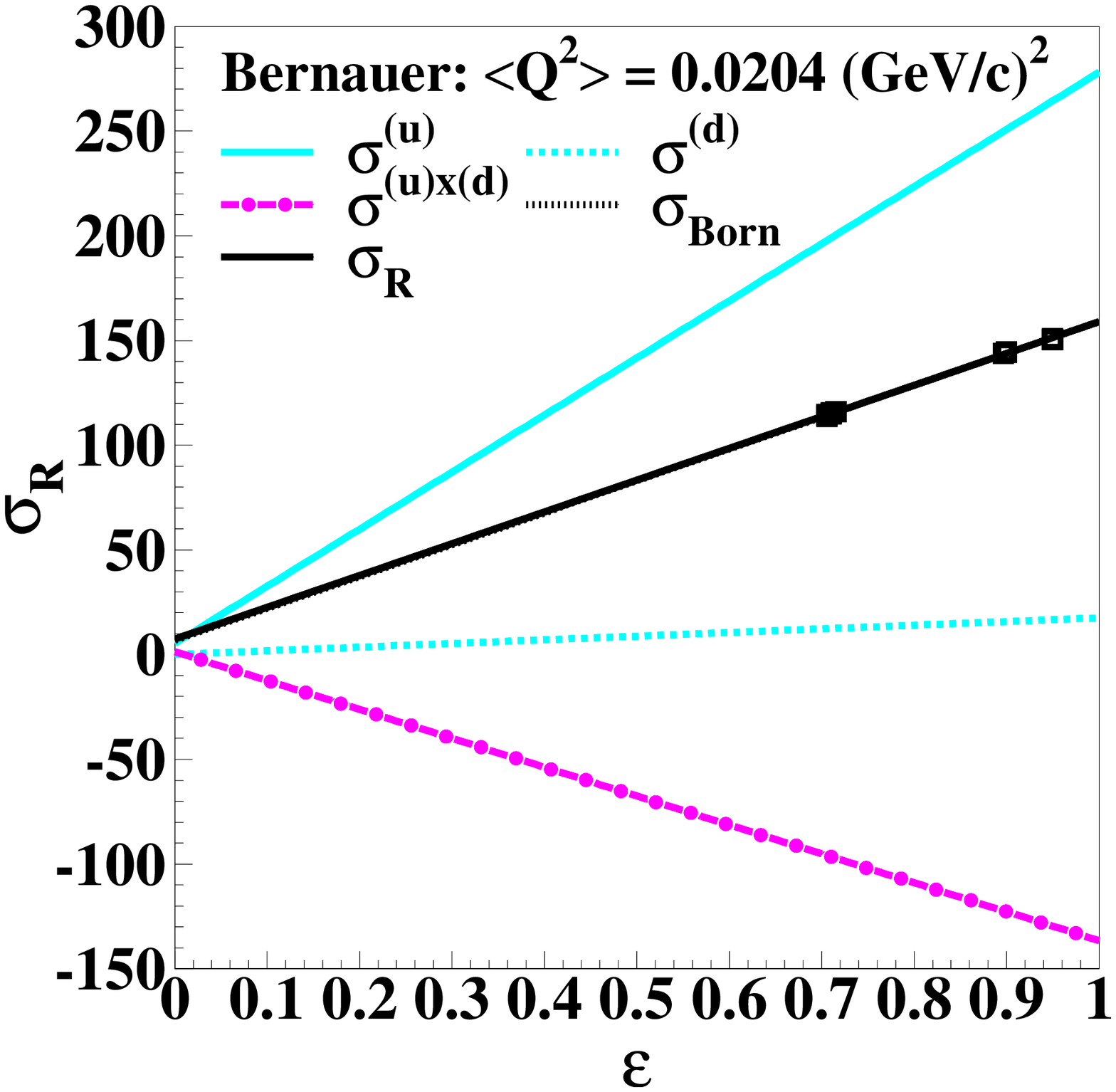} &
\includegraphics*[width=4.3cm]{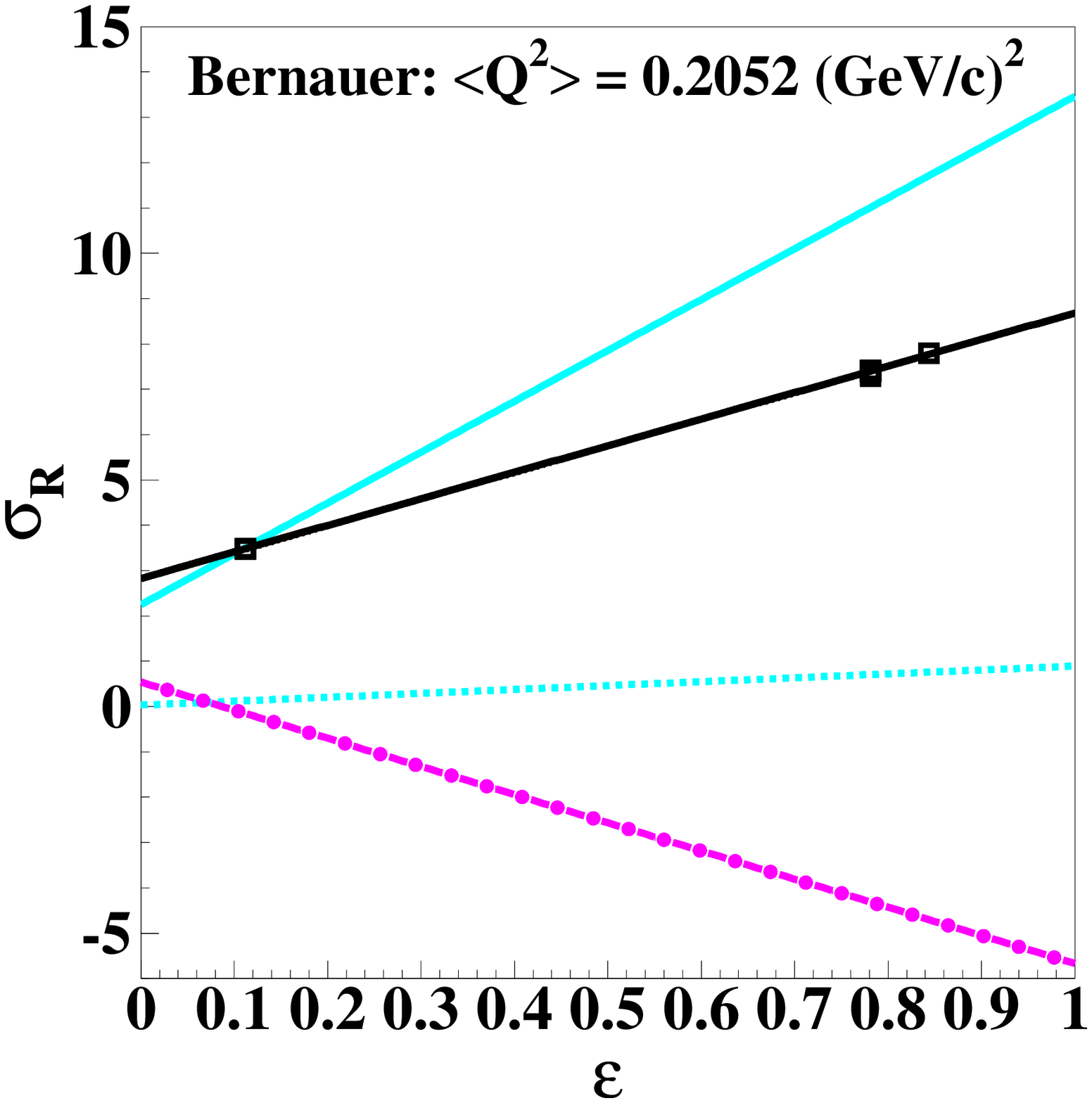} \\
\includegraphics*[width=4.3cm]{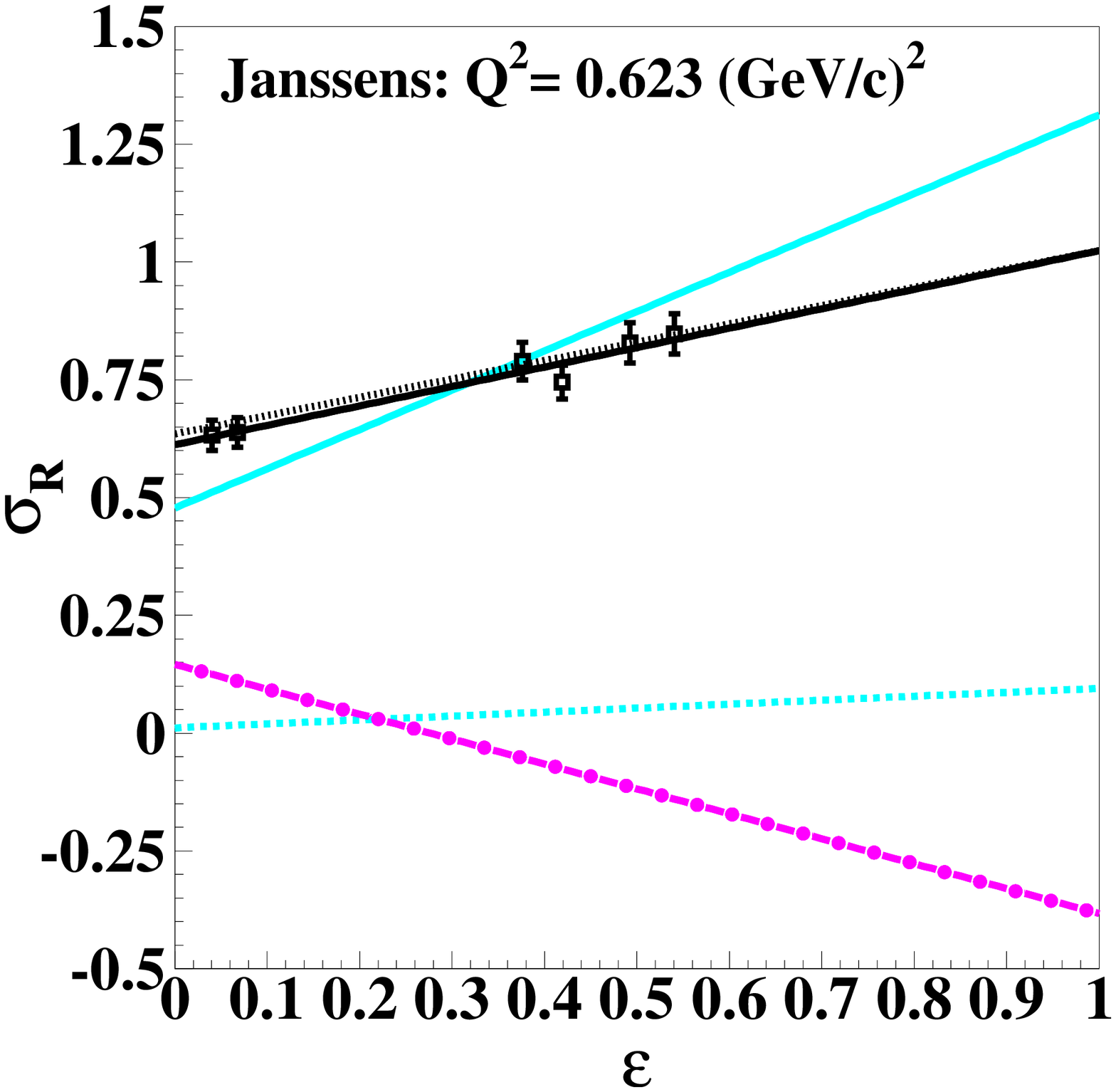} &
\includegraphics*[width=4.3cm]{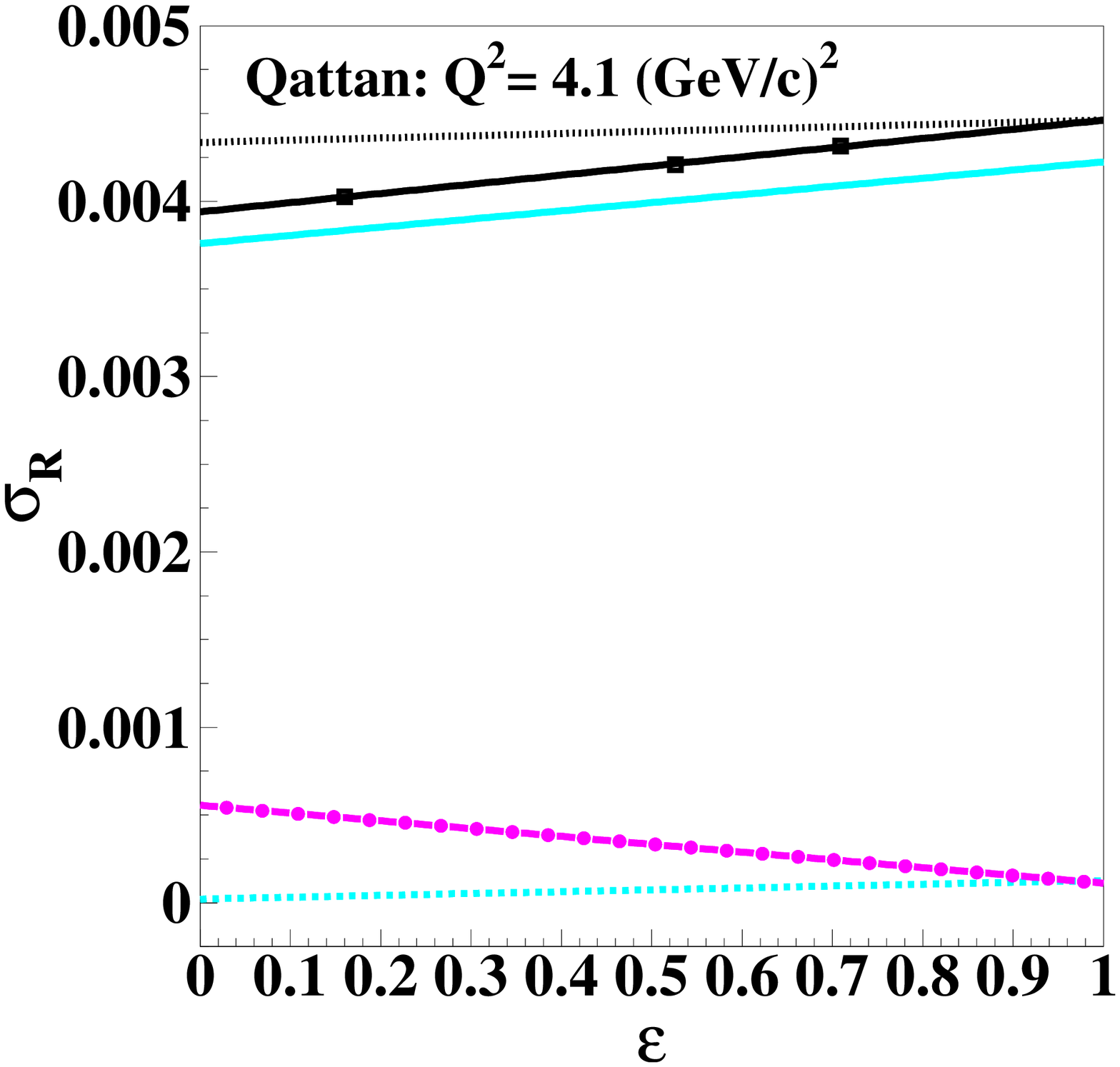} \\
\vspace{-0.5cm}
\end{tabular}
\end{center}
\caption{(Color online) The reduced cross section $\sigma_R$ including TPE (solid line) and without
TPE (dotted black line), along with the flavor-separated contributions to the
Born cross section: $\sigma^{(u)}$ (up-quark), $\sigma^{(d)}$ (down-quark),
and $\sigma^{(u)\times(d)}$ (up-down interference).}
\label{fig:LowQ2SigFlav}
\end{figure}

Figure~\ref{fig:LowQ2SigFlav} shows the total and flavor-separated contributions to the reduced cross section
as a function of $\varepsilon$ for a sample of $Q^2$ values. The down-quark term, $\sigma^{(d)}$ is extremely
small, often smaller than the TPE corrections.  The up-quark $\sigma^{(u)}$, and the up-down quark
interference $\sigma^{(u)\times(d)}$ are the dominant contributions, with the up-down interference term being
comparable in size to the total reduced cross section at low $Q^2$ values.  So while down-quark contribution,
representing the form factor one would obtain if only the down quark distribution was present, is almost
negligible at all values of $\varepsilon$ and $Q^2$, the net impact of the down-quarks on the cross section is
still large, especially at low $Q^2$ values. Note that the strange-quark contributions have been neglected,
and while the contributions coming from $(G_{E,M}^s)^2$ will have a negligible contribution to the cross
section, the up-strange interference term will be significantly larger, and could have a non-negligible 
contribution.

\begin{figure}[!htbp]
\begin{center}
\includegraphics*[width=8.5cm, height=7cm]{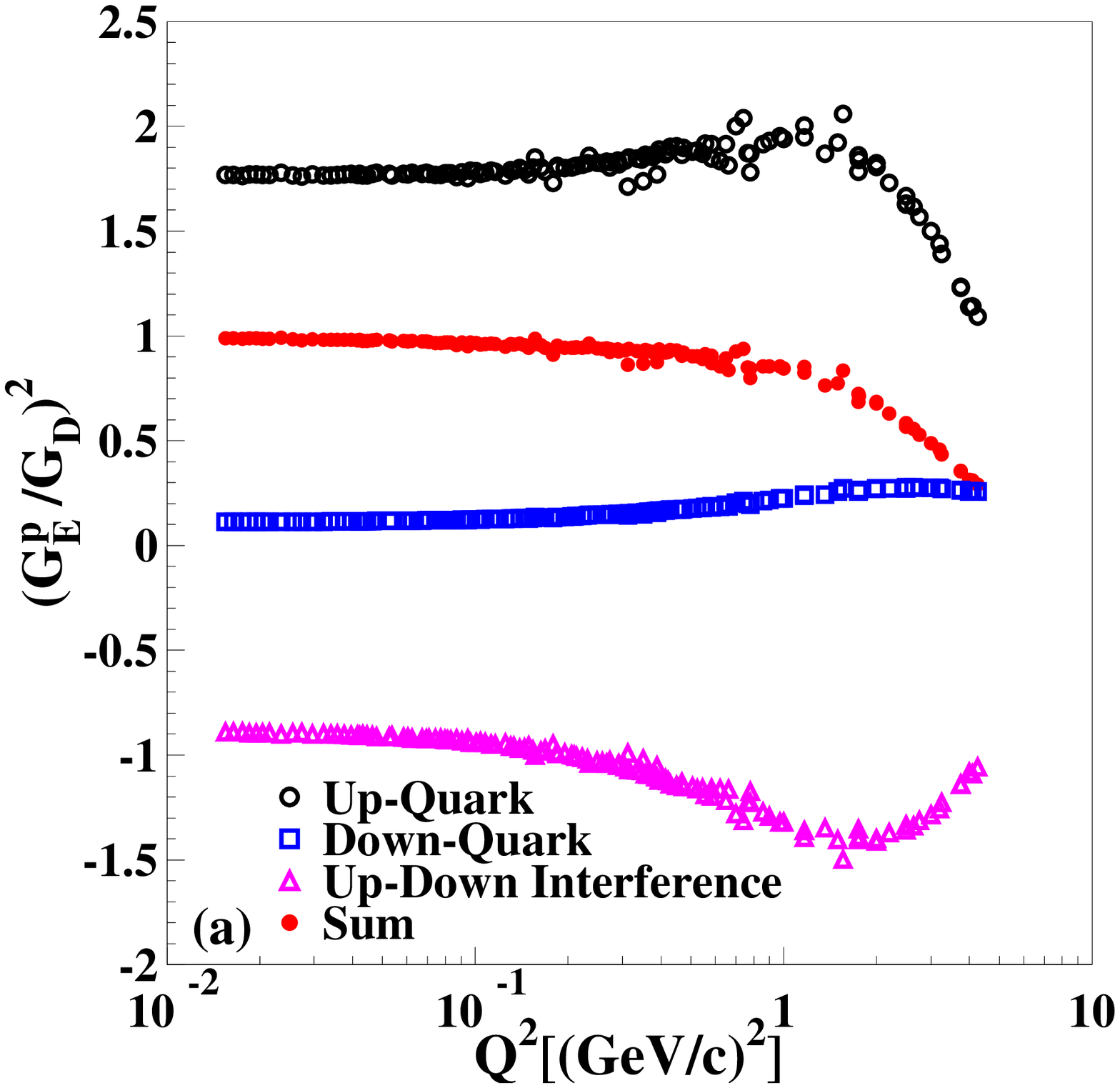}\\
\includegraphics*[width=8.5cm, height=7cm]{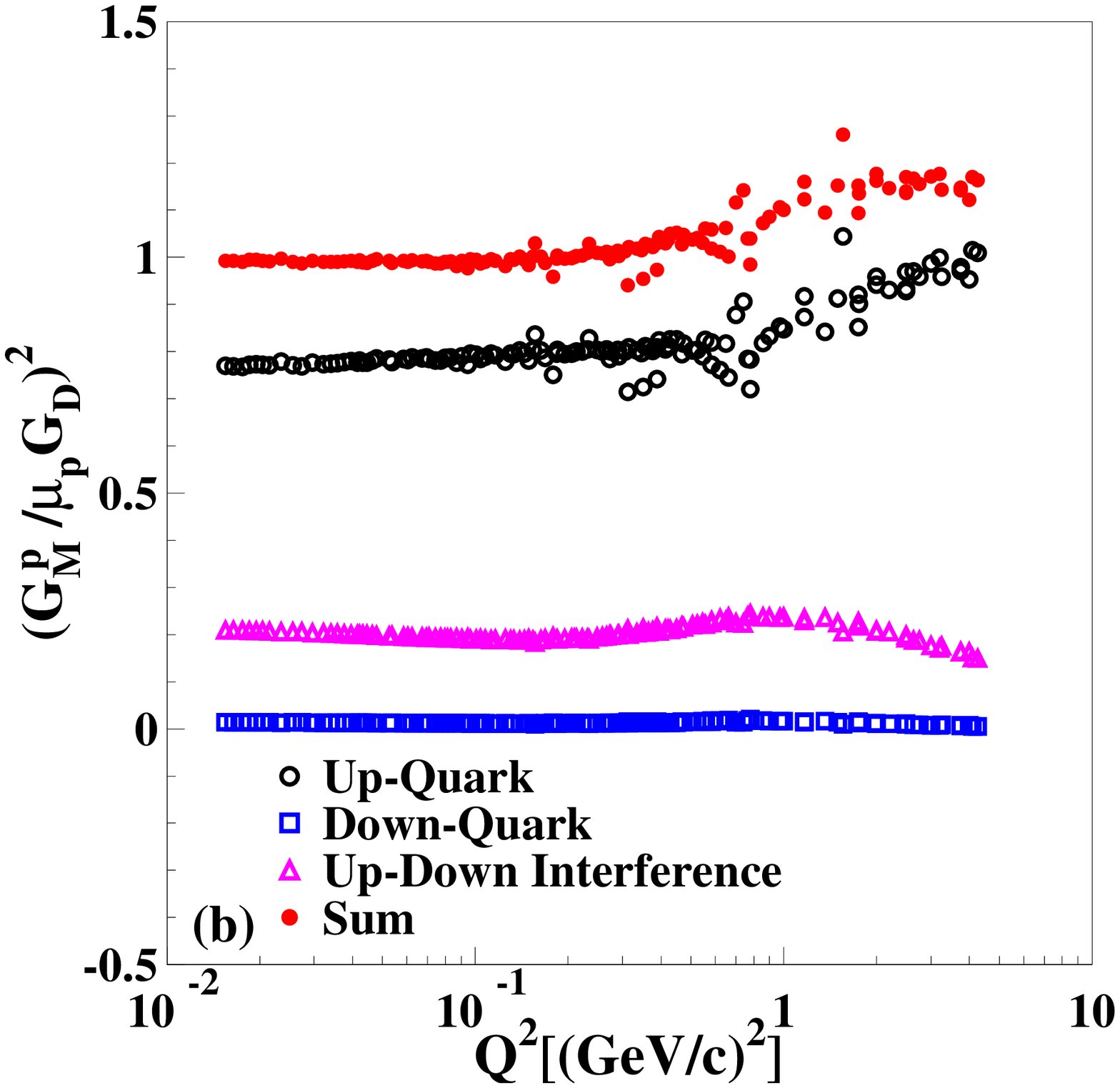}\\
\end{center}
\vspace{-0.5cm}
\caption{(Color online) Flavor-separated contributions to $(G_{E,M}^p)^2$.}
\label{fig:updownFFs}
\end{figure}

To simplify the examination of the flavor-dependent contributions, we examine the contribution of the up, down, 
and interference terms to $(G_E^p)^2$ and $(G_M^p)^2$, as shown in Fig.~\ref{fig:updownFFs}. For $\varepsilon=0$,
the cross section depends on $(G_M^p)^2$ which is dominated by the up-quark contribution, with a roughly 20\%
contribution from the interference term adding to $(G_M^p)^2$ at all $Q^2$ values.

The breakdown of the charge form factor is more complicated.  At low $Q^2$ values, the interference term is roughly
half the size of the up-quark contribution and of the opposite sign, making the interference term nearly identical
in magnitude to the total value of $(G_E^p)^2$. Increasing from $Q^2 \approx 0.1$ to 1 (GeV/c)$^2$, the decrease
in $G_E^p$ is driven by the increase in the magnitude of the (negative) up-down interference term, moderated by a
slight increase in the up-quark contribution.  Above 1-2 (GeV/c)$^2$, the rapid fall of $(G_E^p)^2$, partially
cancelled by an increase in both the down-quark and up-down interference terms, continues the very nearly linear
decrease observed in the $\mugegm$ ratios. Thus, the observed nearly linear behavior comes from a complicated
combination of the up, down, and interference terms, with none of the individual contributions showing such a
monotonically decreasing behavior relative to the dipole form.

The faster falloff of the down-quark contributions was interpreted
in Ref.~\cite{cates11} and references therein as an indication of the possibility
of sizable nonzero strange matrix elements at large $Q^2$ or the
importance of diquark degrees of freedom. While existing measurements
of parity-violating elastic scattering yield very small contributions
from the strange quarks up to $Q^2 \approx 1$~(GeV/c)$^2$~\cite{aniol04,armstrong05,androic10,ahmed12}, 
they still leave open the possibility for significant contributions from  
$G_E^s$ and $G_M^s$ which cancel in the parity-violating observables~\cite{ahmed12,armstrong12},
although there are also results from lattice QCD that the strange-quark 
contribution is small for the charge and magnetic form factors~\cite{paschke11,leinweber05}.

In the diquark model, the singly occurring down-quark in the proton is more
likely to be associated with an axial-vector diquark than a scalar diquark, 
and the contribution of the axial-vector diquark yields a more rapid falloff
of the form factors. The up quarks are generally associated with the more
tightly bound scalar diquarks, yielding a harder form 
factors~\cite{cloet13a, close88, cloet05a, cates11, wilson12, cloet12, gonzalez13, cloet14a,
cloet14b, cloet13c, cloet09, roberts07}. Recent calculations~\cite{cloet14a} suggest that pion 
loop corrections play a crucial rule for $Q^2 \ltorder 1.0$~(GeV/c)$^2$. 
The nucleon form factors were expressed as proton quark sector form factors, 
and was found that the down-quark sector of the Dirac form factor is much softer 
than the up-quark sector as a consequence of the dominance of scalar diquark correlations 
in the proton wave function. On the other hand, the up-quark sector of the Pauli form factor is
slightly softer than in the down-quark sector suggesting that the pion cloud and axial-vector 
diquark correlations dominate the effect of scalar diquark correlations leading to a larger
down-quark anomalous magnetic moment and a form factor in the up-quark sector that is slightly softer
than in the down-quark sector.

\section{Conclusions} \label{conclusions}

In conclusion, we improved on and extended to lower $Q^2$ the previous
extraction~\cite{qattan11a} of form factors and two-photon exchange contributions, as well
as the extraction~\cite{qattan12} of flavor-separated contributions to the proton form
factors. We used new polarization data to obtain an improved parameterization of $\mugegm$
and its uncertainties, and included new cross section data~\cite{bernauer10,janssens65,zhan11} to extend
the analysis to lower $Q^2$ values.

The results for $G_E^p$ are generally in excellent agreement with those extracted based
on a global analysis including calculated TPE hadronic corrections~\cite{arrington07,
venkat11}, as well as some previous phenomenological extractions~\cite{alberico09a, lomon01,
puckett12}. For $G_M^p$, the results disagree noticeably with previous extractions which
applied hadronic TPE corrections and as well as some previous phenomenological extractions.
This is in large part due to the tension between the new low-$Q^2$ Mainz
data~\cite{bernauer14}, although the different approaches for applying TPE corrections have
some impact as well, especially for extractions which neglected TPE.

Our new low $Q^2$ extraction of the TPE contribution yields values of $a(Q^2)$ at the few
percent level. The TPE term shows a change of sign at $Q^2 \approx 0.40$~(GeV/c)$^2$,
which was not seen in our previous extractions and fit~\cite{qattan11a}, but which is
consistent with low $Q^2$ TPE calculations~\cite{arrington04c, blunden05a, borisyuk07, kivel09,
kondratyuk05}. This is the first phenomenological extractions that directly observes this
predicted change of sign at very low $Q^2$ values. This behavior was present in the extraction
of Bernauer~\etal~\cite{bernauer14}, but because the low $Q^2=0$ limit was fixed to the 
Feshbach correction~\cite{mckinley48}, the change of sign is guaranteed by the fit procedure. 

We compared our extracted TPE corrections to previous phenomenological extractions and to
world's data on the ratio of positron--proton and electron--proton scattering cross
sections, $R_{e^{+} e^{-}}$. Our extracted TPE contributions are in generally good agreement with
world's data on $R_{e^{+} e^{-}}$, including recent measurements which show a clear $\varepsilon$
dependence, consistent with the form factor discrepancy, at $Q^2$ values of
1-1.6~(GeV/c)$^2$~\cite{adikaram14, rachek14}.


Inclusion of the new low-$Q^2$ polarization and cross section data allows us to examine the
slope of the form factor at small $Q^2$, which is connected to the flavor contributions to the
RMS charge and magnetization radii of the proton. While the low $Q^2$ data do not allow for a 
precise extraction of these radii, we find that both the up- and down-quark distributions are more 
localized than the overall charge distribution, with $r_d < r_u < r_p$. For the magnetization 
distributions, the up-quark contribution has a smaller radius than the down-quark contribution, 
providing another clear indication that the magnetization distribution does not simply come from 
the spatial distribution of the quarks.

We also express the reduced cross section $\sigma_R$ in terms of these flavor-separated from 
factors to shed light on the contributions of the up-quark, down-quark, and up-down quarks 
interference terms to the Born cross section $\sigma$. The up-quark contribution
$\sigma^{(u)}$ dominates at all $Q^2$ values, while the down-quark contributions
$\sigma^{(d)}$, representing the cross section one would observe if scattering from
only the down quarks, is typically negligible.  The interference term, $\sigma^{(u)\times(d)}$,
is negative and can be sizeable.  While it is always smaller than the up-quark contribution,
it can be comparable in size to the total cross section after the significant cancellation between
the interference terms and the up- and down-quark terms.

\begin{acknowledgments}

This work was supported by Khalifa University of Science, Technology and
Research and by the U.~S. Department of Energy, Office of Science, Office of Nuclear Physics,
under contract DE-AC02-06CH11357. The third author acknowledges the financial support provided by
JUST during his sabbatical leave at the University of Nebraska Omaha, USA.

\end{acknowledgments}

\bibliography{longpaper_TPEparam}

\end{document}